\documentclass{article}

    \PassOptionsToPackage{numbers, compress}{natbib}

 \usepackage[main, final]{neurips_2026}

\usepackage[utf8]{inputenc} 
\usepackage[T1]{fontenc}    
\usepackage{hyperref}       
\usepackage{url}            
\usepackage{booktabs}       
\usepackage{amsfonts}       
\usepackage{nicefrac}       
\usepackage{microtype}      
\usepackage{xcolor}         

\usepackage{amsthm}

\usepackage[T1]{fontenc}    
\usepackage{amsmath}        
\usepackage{makecell}
\usepackage{pifont} 
\newcommand{\cmark}{\textcolor{green!60!black}{\ding{51}}}
\newcommand{\xmark}{\textcolor{red!80!black}{\ding{55}}}

\usepackage[table]{xcolor}  
\usepackage{graphicx}
\usepackage[normalem]{ulem}
\usepackage{algorithm}

\usepackage{algpseudocode}
\usepackage{wrapfig}

\usepackage{xcolor}
\definecolor{s1purple}{RGB}{21, 140, 207} 
\definecolor{s2purple}{RGB}{7, 101, 104} 
\definecolor{kr}{RGB}{165, 0, 52}  
\usepackage{hyperref}
\hypersetup{
  colorlinks=true,
  citecolor=s2purple,
  urlcolor=s1purple,
  linkcolor=kr
}

\title{WaveGen: Truly End-to-End Waveform Generation via Internal Spectral Trajectory Learning}

\author{
    Sang-Hoon Lee$^{1}$\thanks{Equal contribution. Corresponding author: Sang-Hoon Lee (sanghoonlee@ajou.ac.kr)}
    \And
    Ha-Yeong Choi$^{2}$\footnotemark[1]
    \AND
    \\
    $^{1}$Department of Artificial Intelligence, Ajou University, Suwon, Korea\\
    $^{2}$Frontier AI Lab, KT Corp., Seoul, Korea
}
\begin{document}

\maketitle
\begin{abstract}
This work proposes WaveGen, a fully end-to-end diffusion-based text-to-waveform framework based on Spectral Forcing, which jointly learns an internal spectral trajectory and the target waveform trajectory within a single model, enabling direct end-to-end text-to-waveform generation. Unlike recent TTS pipelines, WaveGen does not rely on separately trained or externally decoded intermediate representations, such as neural audio codecs, Mel-spectrogram vocoders, or VAE-based latent models. In addition, WaveGen performs internal text-speech alignment within a single model, eliminating the need for external alignment modules such as duration predictors. To preserve a truly end-to-end formulation, our framework further avoids dependence on self-supervised representations, such as BERT-based text embeddings and Wav2Vec 2.0-based semantic representations, as well as semantic distillation methods for training optimization. To achieve this, we carefully design Spectral Forcing, a model architecture with disentangled diffusion heads, and a training objective based on a multi-scale spectral $v$-loss. We further validate the scalability of end-to-end TTS models by scaling the model size from 0.1B to 0.6B parameters. Using a single architecture, open-source data, and one-stage training, without external modules such as duration predictor or REPA, WaveGen demonstrates fully end-to-end waveform generation achieving state-of-the-art performance on the Emilia training dataset. In particular, WaveGen achieves a WER of 1.47 and a SIM of 0.75 on the Seed-en benchmark. Source code and checkpoints are available at \url{https://github.com/sh-lee-prml/WaveGen}.  
\end{abstract}

\section{Introduction}  
Recent progress in language models and generative models has substantially advanced text-to-speech (TTS) synthesis. 
Large language model (LLM)-based TTS systems \cite{wang2023neural,hu2026qwen3} have replaced conventional acoustic representations with neural audio codecs, enabling speech generation through discrete token modeling and in-context voice prompting. 
In parallel, diffusion and conditional flow matching (CFM)-based TTS models \cite{NEURIPS2023_2d8911db,chen2024f5} have improved the quality and flexibility of Mel-spectrogram generation through masked infilling and zero-shot voice conditioning. 
More recently, latent diffusion approaches \cite{zhou2026hierarchical,peng2026vibevoice} further improve the efficiency of generative modeling by operating in compressed representation spaces rather than directly modeling high-dimensional audio signals.

Despite these advances, most recent TTS systems still rely on intermediate representations and external waveform decoders. 
Codec-based models \cite{copet2023simple,NEURIPS2025_07134952,tian2026ualm} depend on the capacity and quality of neural audio codecs \cite{zeghidour2021soundstream,dfossez2023high,kumar2023high,ji2025wavtokenizer,ye2025codec,NEURIPS2025_220e6dbe,wang2026scaling,della2026beyond}; Mel-spectrogram-based models require separately trained neural vocoders; and latent diffusion models require an additional pre-trained latent.
These intermediate modules have been highly effective in practice, but they also introduce additional training stages, architectural dependencies, and upper-bound constraints on the final quality. 
In particular, high-fidelity waveform generation often requires GAN training, carefully designed architectures, or representation-specific optimization, making the overall pipeline increasingly complex.

Meanwhile, although end-to-end (E2E) TTS models \cite{ren2021fastspeech,kim2021conditional,donahue2021endtoend} were explored in earlier studies, their limited scalability, training instability, and inferior zero-shot quality led the field to converge toward multi-stage pipelines. 
These limitations stem from the high temporal resolution of raw waveforms, which makes text-speech alignment difficult and often requires complex adversarial training for perceptually realistic waveform generation. Only E3-TTS \cite{gao2023e3} demonstrated the potential of waveform-level diffusion for TTS. However, it still relies on BERT-based text representations to facilitate text-waveform alignment, and directly applying diffusion modeling to waveform signals imposes a substantial computational burden due to the high temporal resolution of audio. 

In this work, we propose \textbf{WaveGen}, a fully E2E diffusion-based text-to-waveform framework built upon \textbf{Spectral Forcing}. 
WaveGen directly generates high-resolution waveform from text within a single architecture. The key idea is to revisit the prediction and loss spaces of CFM for raw waveform generation. We found $x$-prediction with $v$-loss is essential for E2E waveform generation.  
Instead of forcing the model to learn waveform dynamics alone, Spectral Forcing jointly guides the model to learn an internal spectral trajectory while optimizing the target waveform trajectory. 
This design provides a structured inductive bias for high-resolution audio generation while preserving a truly E2E text-to-waveform formulation.

To this end, we carefully design WaveGen architecture with disentangled diffusion heads and introduce a multi-scale spectral loss for stable waveform-level flow matching. 
We further investigate the scalability of E2E TTS by increasing the model size from 0.1B to 0.6B parameters, showing that direct text-to-waveform generation can benefit from model scaling when the training objective and architecture are properly designed. 
\vspace{-0.2cm}

Our main contributions are summarized as follows:\vspace{-0.2cm}

\begin{itemize}
    \item We propose \textbf{WaveGen}, a fully E2E text-to-waveform framework that directly generates raw waveform from text without separately trained intermediate representations.
    
    \item We introduce \textbf{Spectral Forcing}, which jointly learns an internal spectral trajectory and the target waveform trajectory, enabling stable CFM-based optimization in the high-resolution waveform domain.
    
    \item We investigate prediction and loss spaces for waveform generation and design a scalable waveform-level diffusion architecture with disentangled diffusion heads.
    
    \item We demonstrate that WaveGen achieves state-of-the-art zero-shot TTS performance using a single architecture, open-source data, and one-stage training, providing evidence that fully E2E text-to-waveform generation can be a viable alternative to increasingly complex multi-stage TTS pipelines.
\end{itemize}
\vspace{-0.3cm}
\section{Related Work} \vspace{-0.2cm}

\subsection{Conventional TTS} \vspace{-0.2cm}

Traditional TTS systems consist of an acoustic model and a neural vocoder. Acoustic models generate acoustic features such as Mel-spectrogram, and neural vocoders convert acoustic features into raw waveform signals. Tacotron \cite{wang17n_interspeech} paved the way of deep-learning based TTS system by introducing attention-based encoder-decoder structure to align text with speech, and Tacotron 2 \cite{shen2018natural} utilized WaveNet \cite{oord2016wavenet} to generate waveform from synthetic Mel-spectrogram. Transformer TTS \cite{li2019neural} verified the effectiveness of encoder-decoder Transformer. However, these models suffer from unstable attention alignment and slow inference speed due to their autoregressive (AR) nature. To address these issues, FastSpeech \cite{ren2019fastspeech} introduced parallel synthesis using a duration which is extracted from teacher AR models. Also, GAN-based neural vocoders accelerated TTS research in terms of both quality and inference speed. Specifically, PWG \cite{yamamoto2020parallel} and MelGAN \cite{kumar2019melgan} introduced GAN-based vocoder, and HiFi-GAN \cite{kong2020hifi} introduced multi-period discriminator (MPD) which significantly enhances perceptual quality of synthetic speech. BigVGAN \cite{lee2023bigvgan} scaled up the neural vocoder via snake periodic activation and an anti-aliasing low-pass filter. iSTFTNet \cite{kaneko2022istftnet} and Vocos \cite{siuzdak2024vocos} enhance the sampling efficiency by modeling features in a low-resolution space with iSTFT.  
\newpage

\begin{table}[t] 
\caption{Comparison of WaveGen with recent TTS systems. WaveGen uniquely achieves a fully E2E single-stage zero-shot TTS without relying on intermediate representations or auxiliary modules. T, C, W, M, L, and B denote text, codec, waveform, Mel, latent, and BERT, respectively.}  \vspace{-0.3cm}
\label{table:summary}
\centering  
\resizebox{1\columnwidth}{!}{
\begin{tabular}{l ccccc|ccc}
    \toprule
 Model  & Qwen3-TTS & VoiceBox & E2/F5-TTS & CosyVoice3 & VoxCPM & VITS& E3-TTS & \textbf{WaveGen} \\
    \midrule
    \textbf{End-to-End (T$\rightarrow$W)} 
        & \xmark & \xmark & \xmark & \xmark & \xmark &\cmark & \xmark & \cmark \\
        
    \textbf{Single-stage} 
        & \xmark & \xmark & \xmark & \xmark & \xmark & \cmark & \xmark & \cmark \\
    \textbf{Zero-shot} 
        & \cmark & \cmark & \cmark & \cmark & \cmark & \xmark  & \cmark & \cmark \\
    \midrule
    \multicolumn{8}{l}{\textit{Dependency on Intermediate Representations (Explicit Waveform Decoder)}} \\
    Codec 
        & \cmark & \xmark & \xmark & \cmark & \xmark & \xmark  & \xmark & \xmark \\
    Vocoder 
        & \xmark & \cmark & \cmark & \cmark & \xmark & \xmark  & \xmark & \xmark \\
    VAE / Latent Model 
        & \xmark & \xmark & \xmark & \xmark & \cmark & \xmark & \xmark & \xmark \\
    GAN-based Training 
        & \cmark & \cmark & \cmark & \cmark & \cmark & \cmark & \xmark & \xmark \\
    \midrule
  \multicolumn{8}{l}{\textit{Dependency on External Alignment Modules}} \\

    Pre-trained LLM 
        & \cmark& \xmark & \xmark & \cmark & \cmark & \xmark & \cmark & \xmark \\
    Duration Predictor
        & \xmark & \cmark & \xmark & \xmark & \xmark & \cmark & \xmark & \xmark \\
    \midrule
    \textbf{Pipeline Complexity} 
        & T$\rightarrow$C$\rightarrow$W &
          T$\rightarrow$M$\rightarrow$W &
          T$\rightarrow$M$\rightarrow$W &
          T$\rightarrow$C$\rightarrow$M$\rightarrow$W &
          T$\rightarrow$L$\rightarrow$W &
          T$\rightarrow$W &
          T$\rightarrow$B$\rightarrow$W &
          \textbf{T$\rightarrow$W} \\
    \bottomrule
\end{tabular}
}  \vspace{-0.6cm}
\end{table}

\paragraph{Neural Codec Language Models} With the emergence of large language models, AudioGen \cite{kreuk2023audiogen} introduces language model-based audio generative models based on text prompt and utilize RVQ-based neural audio codec as a discrete audio representation. Then, VALL-E \cite{wang2023neural} demonstrated neural codec language models can be a strong zero-shot TTS system with large-scale training and speech-level prompting. UniAudio \cite{yang2023uniaudio} and Moshi \cite{defossez2024moshi} proposed efficient RVQ decoding method using multi-scale Transformer. Recently, most models utilize a pre-trained LLM for TTS backbone, and finetuned them to decode semantic token \cite{du2024cosyvoice,du2025cosyvoice} or RVQ tokens \cite{kim2025efficient,hu2026qwen3, gong2026moss} which are extracted from large-scale high-quality audio codec via multi-token prediction (MTP) AR Transformer. Metis \cite{NEURIPS2025_be8987a7} and OmniVoice \cite{zhu2026omnivoice} employ a discrete diffusion language model using bidirectional Transformer.      

\paragraph{Diffusion/CFM}
Grad-TTS \cite{popov2021grad} demonstrated the effectiveness of diffusion models for TTS. VoiceBox \cite{NEURIPS2023_2d8911db} and P-Flow \cite{NEURIPS2023_eb0965da} propose CFM-based TTS models via a masked-infilling training strategy, which significantly improve zero-shot speech synthesis through speech prompting. E2-TTS \cite{eskimez2024e2} introduces novel text-speech alignment learning with filler tokens. F5-TTS \cite{chen2024f5} and ZipVoice \cite{zhu2025zipvoice} introduce efficient CFM-based TTS models without any external duration predictor. Also, hybrid models such as CosyVoice \cite{du2024cosyvoice,du2025cosyvoice} utilized CFM-based Mel-spectrogram decoder conditioned on semantic tokens. Meanwhile, latent diffusion models (LDM) \cite{lee2024ditto} have been investigated for efficient TTS modeling. Furthermore, continuous language models such as VoxCPM \cite{zhou2026hierarchical} and VibeVoice \cite{peng2026vibevoice} show their effectiveness of AR diffusion via highly-compressed latent representation.

\subsection{End-to-End TTS} 

E2E TTS has long been investigated as a way to simplify conventional cascaded speech synthesis pipelines. Early neural TTS systems such as Tacotron \cite{wang17n_interspeech} established a unified sequence-to-sequence framework for text-to-acoustic modeling, but still required a Griffin-Lim algorithm \cite{griffin1984signal}. ClariNet \cite{ping2018clarinet} explored parallel waveform generation through knowledge distillation from autoregressive WaveNet models. FastSpeech 2s \cite{ren2021fastspeech} extended non-autoregressive TTS toward direct waveform generation. EATS \cite{donahue2021endtoend} jointly optimized learnable duration modeling and waveform generation with adversarial training. VITS \cite{kim2021conditional} introduced a strong text-to-waveform framework by combining variational inference, normalizing flows \cite{kim2020glow}, and adversarial waveform training \cite{kong2020hifi}. 

Despite these advances, prior E2E TTS systems still have important limitations: many rely on adversarial training to obtain perceptually realistic waveforms, which require carefully designed multiple discriminators, auxiliary losses, and significantly long training. More importantly, most of these models were not primarily designed for zero-shot speech synthesis, where the model must generate high-quality speech for unseen speakers from text and speech prompts. As a result, recent zero-shot TTS research has largely shifted toward codec-based language models, Mel-spectrogram diffusion models, or LDM rather than direct waveform-level generation.

Recently, E3-TTS \cite{gao2023e3} revisited diffusion-based text-to-waveform generation, demonstrating the potential of waveform-level generative modeling for TTS; however, it still relies on pretrained BERT-based text representations to facilitate text-waveform alignment, and directly applying diffusion modeling to high-resolution waveform signals introduces substantial computational cost. In contrast, this paper revisits fully E2E TTS under a stricter formulation: a single model directly generates waveform from text without neural audio codecs, Mel-spectrogram decoders, VAE, neural vocoders, duration predictors, self-supervised representations, or adversarial training as indicated in Table \ref{table:summary}.

\begin{figure*}[t]
    \centering
    {\includegraphics[width=1\textwidth]{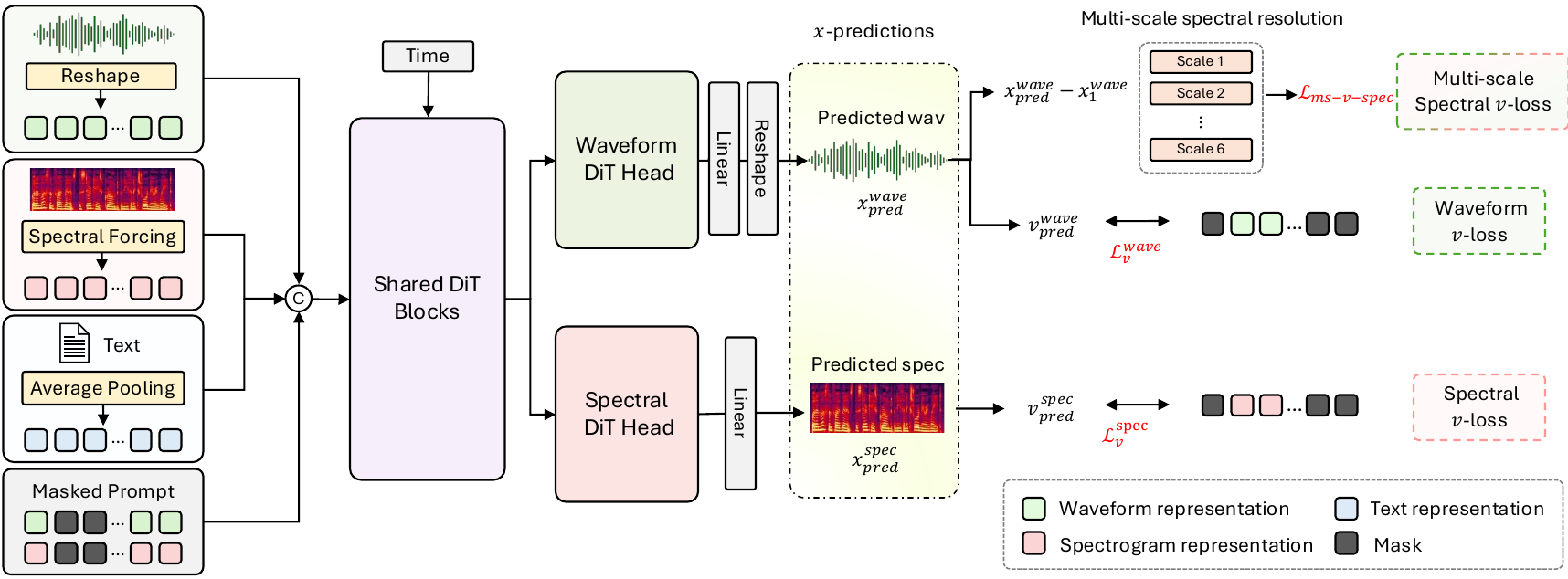}} \vspace{-0.6cm}
    \caption{E2E text-to-waveform diffusion training details. We reshape high-resolution waveform signals (24,000 Hz) into a low-resolution space (100 Hz) for efficient E2E waveform diffusion. Then, we introduce Spectral Forcing for internal spectral guidance and simply concatenate average-upsampled text embeddings. We use shared DiT blocks for text-speech alignment and disentangled DiT heads for representation refinement. We adopt $x$-prediction as the prediction space, and $v$-loss as the target loss space, and introduce a new multi-scale $v$-space spectral loss} \vspace{-0.5cm}
    \label{fig1:overall}
\end{figure*}     
\section{WaveGen}  \vspace{-0.3cm} 

In this paper, we propose a fully E2E single-stage diffusion-based text-to-waveform generative model as illustrated in Figure \ref{fig1:overall}. We analyze raw waveform as generative targets (Section \ref{section3_1}), introduce a novel framework (Section \ref{section3_2}), and efficient architecture (Section \ref{section3_3}). Also, we revisit the prediction space and loss space for raw waveform (Section \ref{section3_4}) and introduce the improved loss (Section \ref{section3_5}). 
\vspace{-0.3cm}
\subsection{CFM-based Waveform Generation} \label{section3_1} \vspace{-0.2cm}

Recently, diffusion and CFM-based waveform generation models \cite{roman2023from,welker2025flowdec,lee2024accelerating,lee-choi-2026-hierarchical} have emerged as a promising alternative to GAN-based models, offering a new direction for high-fidelity speech synthesis. PeriodWave \cite{lee2025periodwave} introduced a CFM-based periodic-aware generator for waveform generation. RFWave \cite{liu2025rfwave} presented efficient CFM-based models with STFT/iSTFT. Flow2GAN \cite{yao2026flowgan} improved CFM objective of waveform in spectral space. StreamFlow \cite{NEURIPS2025_07134952} proposed streaming flow matching for streaming waveform generation and efficient waveform diffusion Transformers using linear-reshape transformation. However, these works only employ waveform-level diffusion with strongly aligned acoustic conditions such as Mel-spectrogram or pre-trained neural audio codec. In our study, basic CFMs fail to align text-speech in high-resolution waveform space without any external modules. 
\vspace{-0.3cm}
\subsection{Spectral Forcing} \label{section3_2} \vspace{-0.2cm}
Due to the difficulty of waveform-level generation, previous works utilize intermediate representations such as Mel-spectrogram. To bridge the gap  between text and waveform, we introduce Spectral Forcing, a method to jointly learn an internal spectral trajectory and the target waveform trajectory, enabling stable CFM-based optimization in the high-resolution waveform space. 
Specifically, we introduce a low-resolution spectral trajectory as an auxiliary generation target, which provides a structured inductive bias for text-speech alignment. Unlike cascaded approaches, the spectral trajectory is jointly generated within the same model and is used only as internal frequency-rich guidance, rather than as an external intermediate representation.

We adopt a simple DiT \cite{peebles2023scalable}-only architecture with AdaLN-SOLA \cite{hai25_interspeech}. The model takes as input the concatenation of the reshaped noisy waveform $x^{wave}_t$ and its noisy spectral representation $x^{spec}_t$ which is a lower-resolution Mel-spectrogram with 64 bins. For text conditioning, we adopt average pooling of ZipVoice \cite{zhu2025zipvoice,zhu2025zipvoicedia} that expands the average-pooled text embeddings along the temporal dimension to match the speech length without any external encoder. Following \cite{NEURIPS2023_2d8911db, chen2024f5}, we utilize masked-infilling training for prompted generation by concatenating masked representations together. 
 
\vspace{-0.3cm}
\subsection{Disentangled DiT Heads} \label{section3_3} \vspace{-0.3cm}
We found that concatenating input representations is necessary for better text-speech alignment, but disentangling DiT heads for each representation is required for robust estimation. We use shared DiT blocks as an alignment backbone, and simply add shallow disentangled DiT heads as representation-specific generation. Baseline model uses 14 shared DiT blocks and two disentangled heads. We only compute the loss on the masked regions of the target representation.
 
\subsection{Decoupled Prediction and Loss Spaces for Raw Waveform Generation} \label{section3_4}\vspace{-0.2cm}
Recent flow matching-based waveform reconstruction models, such as PeriodWave \cite{lee2025periodwave} and RFWave \cite{liu2025rfwave}, have demonstrated the potential of flow-based models for raw waveform reconstruction by improving periodic modeling and sampling efficiency. Despite these advances, we observe that conventional velocity-based objectives can be insufficient for preserving fine-grained waveform energy, often leading to degraded perceptual quality in generated audio.

Motivated by JiT \cite{li2025back}, which highlights the effectiveness of clean-data prediction for high-dimensional raw signals, and Latent Forcing \cite{baade2026latent} and ReGen \cite{lee2026regen}, which leverage latent-assisted generation trajectories to better structure raw-space generation, we revisit the design of prediction and loss spaces in raw waveform CFM. 
Specifically, our model predicts clean waveform and spectral targets through Waveform DiT and Spectral DiT, while computing the training objective in the velocity space. This decoupled $x$-prediction and $v$-loss strategy enables effective waveform energy modeling without introducing the severe audible artifacts observed with direct $x$-loss optimization. Following the standard flow interpolation, we construct the intermediate sample in each domain $d \in \{\mathrm{wave}, \mathrm{spec}\}$ as $x_t^d=(1-t)x_0^d+t x_1^d$, where $x_0^d$ is Gaussian noise and $x_1^d$ is the clean target. Instead of directly predicting the velocity field, the model predicts the clean target $x_{pred}^d$. We then convert this prediction into the velocity space as:
\vspace{-0.1cm}
\begin{equation}
v_{pred}^d = \frac{x_{pred}^d - x_t^d}{1 - t},
\qquad v_t^d = \frac{x_1^d - x_t^d}{1-t}.
\end{equation} \vspace{-0.1cm}
The loss is computed only on the masked generation region:
\begin{equation}
\mathcal{L}_v^d = \frac{\left\|\mathcal{M}^d \odot \left(v_{pred}^d - v_t^d\right)\right\|_2^2}{\left\|\mathcal{M}^d\right\|_1},
\qquad
d \in \{\mathrm{wave}, \mathrm{spec}\},
\end{equation}
where $\mathcal{M}^d$ denotes the domain-specific training mask and $\odot$ is element-wise multiplication. This yields the waveform velocity loss $\mathcal{L}_v^{wave}$ and the spectral velocity loss $\mathcal{L}_v^{spec}$. 
\vspace{-0.1cm}
\subsection{Multi-Scale Energy-Normalized Spectral $v$-Loss} \label{section3_5}\vspace{-0.1cm}
 
Flow2GAN \cite{yao2026flowgan} introduces spectral energy-adaptive loss scaling to alleviate the loss-perception mismatch in flow matching. Following this insight, we adopt energy-normalized spectral supervision for the CFM velocity objective and extend it to multiple spectral resolutions, enabling the model to capture time-frequency structures at different resolutions.
As described in Section 3.4, our model adopts $x$-prediction while computing the objective in the velocity space. We define the waveform-domain velocity difference as:
\begin{equation}
 \Delta_v^{wave} = x_{pred}^{wave} - x_1^{wave} = (1-t)\left(v_{pred}^{wave}-v_t^{wave}\right). 
\end{equation} 
We use $\Delta_v^{wave}$ for spectral supervision. For each scale $k \in \mathcal{K}$, we first compute a power spectrogram and apply a linear filterbank projection: 
\begin{equation}
\Phi_k(x) = \left| \mathrm{STFT}_k(x) \right|^2 \mathbf{F}_k,
\end{equation} 
where $\mathbf{F}_k$ denotes the linear filterbank matrix at the $k$-th spectral resolution. The proposed multi-scale spectral $v$-loss is then defined as:
\begin{equation}
\mathcal{L}_{ms\text{-}v\text{-}spec}
= \sum_{k \in \mathcal{K}} \frac{ \left\| \mathcal{M}_k^{wave} \odot \Phi_k(\Delta_v^{wave})
\odot \left( \Phi_k(x_1^{wave}) + \epsilon \right)^{-\frac{1}{2}} \right\|_1}{\left\|\mathcal{M}_k^{wave}\right\|_1 },
\end{equation} 
where $\mathcal{M}_k^{wave}$ is the mask for the $k$-th spectral resolution. Here, $\mathcal{K}$ is a set of spectral configurations with $n_{\mathrm{bin}}=[512,256,128,64,32,16]$,
$n_{\mathrm{fft}}=w=[1920,960,480,240,120,60]$, and $h=[480,240,120,60,30,15]$, where $w$ and $h$ denote the window size and hop size, respectively. The linear filter projection decomposes wave-domain velocity difference across frequency bands, and the ground-truth energy normalization prevents high-energy regions from dominating the loss. We follow Flow2GAN \cite{yao2026flowgan} for the scaling hyperparameters, using $\epsilon=10^{-7}$ and clamping the scale to $[0.01, 100]$.  
The final objective is denoted as:
\begin{equation}
\mathcal{L}_{total} = \mathcal{L}_v^{wave} + \lambda_{spec} \mathcal{L}_v^{spec} + \lambda_{ms} \mathcal{L}_{ms\text{-}v\text{-}spec},
\end{equation} 
where $\lambda_{spec} = 0.1$ and $\lambda_{ms} = 0.1$.

\section{Experiment and Result} \label{sec4} \vspace{-0.2cm}
\subsection{Experimental Setup} \vspace{-0.2cm}
\paragraph{Dataset} For small-scale TTS and ablation studies, we utilized LibriTTS (CC-BY-4.0) \cite{zen19_interspeech}. LibriTTS is a high-quality speech dataset with a sampling rate of 24,000 Hz in WAV format. We filtered LibriTTS to include only utterances with durations of up to 12s for efficient training (319,627 samples). For large-scale training, WaveGen-S/B (v0.1) use the English subsets of Emilia (CC-BY-NC-4.0) and Emilia-YODAS (CC-BY-4.0) \cite{he2025emilia} provided as MP3-compressed audio with a sampling rate of 24,000 Hz. Following \cite{chen2024f5}, we filter out several speakers and use utterances with durations of up to 30s. We also released WaveGen (0.6B) trained with multilingual Emilia datasets.  

\vspace{-0.2cm}
\paragraph{Training} We train LibriTTS models with a learning rate of 1$\times$10$^{-4}$ using the AdamW optimizer \cite{loshchilov2018decoupled}, a fixed batch size of 128 for 1.1M steps. We train Emilia models with a learning rate of 1$\times$10$^{-4}$ using AdamW optimizer \cite{loshchilov2018decoupled}, a dynamic batch size with total 480s frame of waveform for 1.1M steps. All models except for the 0.6B models are trained on four GPUs using mixed precision. Due to limited computational resources, we train the models only for four epochs. We train WaveGen (0.6B) on eight GPUs for 11 epochs. 
During training, we use a masked-infilling strategy by dropping the speech prompt and text condition (0.3 and 0.2, respectively).

\vspace{-0.2cm}
\paragraph{Sampling}
We adopt the Euler method as the ODE solver. For CFG scheduling, we build upon the strategy used in ZipVoice, where the text condition is dropped only at early time steps, and all ablation studies are conducted using this. However, when using a threshold of $t<0.5$, we found that this substantially degrades speaker similarity in our model. Therefore, we restrict text-condition dropout to $t<0.02$. We fix NFE to 32 and CFG to 2 in our experiments.

\begin{table}[t]
    \centering
    \caption{Zero-shot TTS results on Seed-en Benchmark. Note that parameter size (Params) of baseline models only indicates their TTS module, but they require additional modules for waveform decoding. } \vspace{-0.2cm}
    \label{tab:tts_seed_benchmark}
    \resizebox{1\linewidth}{!}{
    \begin{tabular}{lc|cc|ccccccc|cc}
        \toprule 

        Model & E2E  & Params & Data & Vocoder & Codec & VAE & SPK Enc. & LLM & GAN & Diffusion & \textbf{WER} $\downarrow$ & \textbf{SIM} $\uparrow$ \\
        \midrule
        GT& - & - & - & -& -& -& -& -& - & - & 2.14 & 0.73\\
        \midrule
        SparkTTS \cite{wang2025spark} & \xmark  & 0.5B & 100k & \xmark & \cmark & \xmark & \cmark & \cmark & \cmark &\xmark & 1.98 & 0.58\\  
        TadiCodec-AR  \cite{NEURIPS2025_d8a12fde} & \xmark  & 4B & 100k  & \cmark & \cmark & \xmark & \xmark & \cmark & \cmark & \cmark & 2.28 & 0.65 \\
        \midrule
        F5-TTS \cite{chen2024f5} & \xmark & 0.3B & 100k & \cmark & \xmark & \xmark & \xmark & \xmark & \cmark & \cmark   & 2.00 & 0.67  \\
        ZipVoice \cite{zhu2025zipvoice}& \xmark& 0.1B & 100k  & \cmark & \xmark & \xmark & \xmark & \xmark & \cmark & \cmark  & 1.70 & 0.69  \\
        CosyVoice2 \cite{du2024cosyvoice}& \xmark & 0.5B & 166k  & \cmark & \cmark & \xmark & \cmark & \cmark & \cmark & \cmark  & 3.09 & 0.65  \\
        CosyVoice3-RL \cite{du2025cosyvoice}& \xmark  & 0.5B & -  & \cmark & \cmark & \xmark & \cmark & \cmark & \cmark& \cmark  & 1.68 & 0.69  \\
        \midrule
        Qwen3-TTS \cite{hu2026qwen3} & \xmark & 0.6B & -  & \cmark & \cmark & \xmark & \cmark & \cmark & \cmark& \xmark  & 1.68 & 0.70  \\
        Qwen3-TTS \cite{hu2026qwen3} & \xmark & 1.7B & -  & \cmark & \cmark & \xmark & \cmark & \cmark & \cmark& \xmark  & 1.50 & 0.71  \\
        MOSS-TTS \cite{gong2026moss} & \xmark & 1.7B & -  & \cmark & \cmark & \xmark & \cmark & \cmark & \cmark& \xmark  & 1.93 & 0.73  \\
       \midrule
        VibeVoice \cite{peng2025vibevoice} & \xmark & 1.5B & -  & \xmark & \xmark & \cmark & \xmark & \cmark & \cmark & \cmark  & 3.04 & 0.68  \\
        VibeVoice-RT \cite{peng2025vibevoice} & \xmark & 0.5B & - & \xmark & \xmark & \cmark & \xmark & \cmark & \cmark & \cmark & 2.05 & 0.63  \\
        VoxCPM-Emilia \cite{zhou2026hierarchical} & \xmark & 0.5B & 100k  & \xmark & \xmark & \cmark & \xmark & \cmark & \cmark & \cmark & 2.34 & 0.68 \\
        VoxCPM \cite{zhou2026hierarchical} & \xmark & 0.5B & -  & \xmark & \xmark & \cmark & \xmark & \cmark & \cmark & \cmark & 1.85 & 0.72 \\
        \midrule
        WaveGen-S  & \cmark & 0.1B & 120k& \xmark & \xmark & \xmark & \xmark & \xmark & \xmark & \cmark & 4.42& 0.53 \\ 
        WaveGen-B  & \cmark & 0.4B & 120k& \xmark & \xmark & \xmark & \xmark & \xmark & \xmark & \cmark & 1.77&0.63 \\ 
        WaveGen & \cmark & 0.6B & 100k& \xmark & \xmark & \xmark & \xmark & \xmark & \xmark & \cmark & 1.47& 0.75 \\ 
        \bottomrule
    \end{tabular}
    } \vspace{-0.3cm}
\end{table}
\begin{table}[t]
    \centering
    \caption{Subjective evaluation results on Seed-en benchmark} \vspace{-0.2cm}
    \label{tab:tts_seed_benchmark_mos}
    \resizebox{1\linewidth}{!}{
    \begin{tabular}{l|ccccccc}
        \toprule 
        Model &GT & F5-TTS & CosyVoice3 & Qwen3-TTS-0.6B & Qwen3-TTS-1.7B & WaveGen-S & WaveGen-B \\
        \midrule
        nMOS & 4.117$\pm$0.112 & 3.938$\pm$0.119 & 3.917$\pm$0.128 & 3.883$\pm$0.123 & 3.954$\pm$0.117 & 3.958$\pm$0.129 & 4.000$\pm$0.126 \\
        sMOS & 3.932$\pm$0.038 & 3.989$\pm$0.037 & 3.972$\pm$0.038 & 3.968$\pm$0.037 & 4.008$\pm$0.037 & 3.952$\pm$0.037 & 3.988$\pm$0.037  \\

        \bottomrule
    \end{tabular}
    } \vspace{-0.5cm}
\end{table}

\subsection{Zero-shot TTS results} 

We evaluate zero-shot TTS performance on the Seed-en benchmark dataset \cite{anastassiou2024seed}. We compute WER and SIM to assess speech intelligibility and speaker similarity using Whisper and WavLM-Large, respectively. Table \ref{tab:tts_seed_benchmark} demonstrates the effectiveness of E2E TTS training without relying on external modules, and shows the scalability of the proposed models by achieving promising performance compared to large-scale TTS systems that require multi-stage training. 

We also conduct a subjective evaluation using Amazon Mechanical Turk (MTurk) with details provided in Appendix \ref{appendix:mturk}. We compare WaveGen with publicly available strong baselines, including F5-TTS \cite{chen2024f5}, CosyVoice3 \cite{du2025cosyvoice}, and Qwen3-TTS \cite{hu2026qwen3}. Notably, our model achieves comparable mean opinion scores (MOS) for both naturalness (nMOS) and similarity (sMOS) as reported in Table \ref{tab:tts_seed_benchmark_mos}. We encourage readers to listen to the audio samples on our demo page. Additionally, we provide extensive ablation studies in the following subsections.

\newpage
\subsection{Spectral Forcing}\vspace{-0.25cm}
\begin{wraptable}{r}{7.5cm} \vspace{-0.6cm}
  \caption{Comparison of Spectral Forcing}
  \label{table:spectralforcing}
  \centering
 \resizebox{7.5cm}{!}{
  \begin{tabular}{lc|ccc|c}
    \toprule
    Type & Bin &  WER & SIM & DNSMOS & Analysis   \\
   \midrule
    No SF & - & 123.67&0.19& 3.20 & Fails to align\\
   \midrule
    Mel & \textcolor{red!80!black}{256} & 126.50& 0.35 & 3.17 &  \textcolor{red!80!black}{Denoising}\\
    Mel & \textcolor{red!80!black}{128} & 18.67 & 0.52 & 3.55 & \textcolor{red!80!black}{Reconstruction} \\
    Mel & 80 & 6.80&0.50 & 3.62 & Mel Bins \textcolor{green!80!black}{$\downarrow$} \textcolor{red!80!black}{$\uparrow$}\\
   Mel & \textcolor{green!60!black}{64} & 4.96&0.47&3.72 & \textcolor{green!60!black}{Faster}\\
    Mel &  \textcolor{green!60!black}{32} & 4.26&0.48&3.64 & \textcolor{green!60!black}{Convergence}   \\
   \midrule
   Linear & 481 & & NaN & \\
    Linear & 128 & & NaN & & Unstable\\
    Linear & 64 & & NaN & \\

    \bottomrule
  \end{tabular}
  } \vspace{-0.5cm}
\end{wraptable} 

We found that learning the waveform trajectory alone does not guarantee text-speech alignment. To bridge the gap between text and speech, we introduce Spectral Forcing, which jointly learns an internal spectral trajectory within a single model.
We first use a linear spectrogram as the target spectral representation, but it leads to unstable training. We therefore adopt the Mel-spectrogram as a target representation and analyze the effect of the number of Mel bins in Table \ref{table:spectralforcing}. Larger Mel bins make the model focus more on waveform reconstruction rather than text-speech alignment. In contrast, smaller Mel bins provide a stronger information bottleneck, encouraging earlier alignment and faster convergence. We adopt 64 Mel bins, which offers a favorable trade-off between alignment, convergence, and high-frequency detail generation.

\vspace{-0.25cm}
\subsection{End-to-End Waveform Generation}\vspace{-0.25cm}
\begin{wraptable}{r}{8cm} \vspace{-0.7cm}
  \caption{Comparison of E2E and cascaded models}
  \label{table:e2e}
  \centering
 \resizebox{8cm}{!}{
  \begin{tabular}{lccc|ccc}
    \toprule
    Method & Params.  & NFE & RTF &  WER & SIM & DNSMOS    \\
   \midrule
    End-to-End & 407M & 32 &0.24 & 2.91 & 0.53& 3.67\\
    End-to-End & 210M & 32 &0.12 & 4.68 & 0.50 & 3.71 \\
    End-to-End & 113M & 32 & 0.08 &  4.96 & 0.47 & 3.72 \\
    \midrule
    Vocoding (GT Mel) & 113M & 32 & 0.08 & 1.47&0.70&3.32\\
    Vocoding (GT Mel) & 113M & 16 & 0.04 & 1.57& 0.70 & 3.33 \\
    \midrule
    Two-Stage & 226M & 32 + 32 & 0.16 & 4.39 & 0.46 & 3.53 \\
    Two-Stage & 226M & 16 + 16 & 0.09& 5.17 & 0.47 & 3.59\\
    \bottomrule
  \end{tabular}
  } \vspace{-0.5cm}
\end{wraptable}

To evaluate the effectiveness of E2E training, we train a two-stage TTS system consisting of a text-to-Mel model and a vocoder. For both stages, we use the same DiT architecture with masked-infilling training.
Table \ref{table:e2e} shows that the vocoder can be successfully trained even without GAN-based training. However, the two-stage pipeline still suffers from the inherent limitations of cascaded generation, where the intermediate Mel representation constrains the waveform generation process and errors from the acoustic model can propagate to the vocoder.
In contrast, E2E models directly optimize text-to-waveform generation within a single model, achieving better WER, SIM, and DNSMOS under the same NFEs (total NFE of 32). These results demonstrate the advantage of E2E over cascaded TTS systems. We also report the real-time factor (RTF) on NVIDIA A100 without inference optimization.

\vspace{-0.25cm}
\subsection{Architecture Design}\vspace{-0.25cm}
\begin{wraptable}{r}{6cm} \vspace{-0.65cm}
  \caption{Disentangled DiT Heads}
  \label{table:ddt}
  \centering
 \resizebox{6cm}{!}{
  \begin{tabular}{cc|ccc}
    \toprule
    \makecell{ $\#$Shared \\ DiT} &  \makecell{$\#$ Disentangled \\DiT} &  WER & SIM & DNSMOS    \\
   \midrule
    18 & 0 & 7.83& 0.44& 3.60\\
    14 & 2 (4) &  4.96 & 0.47 & 3.72 \\
    10 & 4 (8) & 7.64& 0.46& 3.65 \\
    \bottomrule
  \end{tabular}
  } \vspace{-0.5cm}
\end{wraptable}
We compare the effectiveness of disentangled DiT heads for multi-representation generation. Using only shared DiT blocks shows poor performance, suggesting that a fully shared architecture is insufficient for generating multiple representations. We therefore add shallow disentangled DiT heads on top of shared DiT blocks, enabling the model to maintain a shared alignment backbone while learning representation-specific refinement. Table \ref{table:ddt} shows that this design substantially improves WER, SIM, and DNSMOS. However, using too many disentangled blocks degrades performance due to degraded alignment. We therefore adopt 14 shared DiT blocks with 2 disentangled DiT blocks.

\vspace{-0.25cm}
\subsection{Decoupled Prediction and Loss Spaces}\vspace{-0.25cm}
\begin{wraptable}{r}{6.0cm} \vspace{-0.7cm}
  \caption{Comparison of prediction space}
  \label{table:loss}
  \centering
 \resizebox{6.0cm}{!}{
  \begin{tabular}{ll|l|ccc}
    \toprule
     Pred &  Loss & SF &  WER & SIM & DNSMOS    \\
   \midrule
    $x$-pred & $v$-loss &\xmark& 123.67&0.19&3.20 \\
    $v$-pred & $v$-loss &\xmark & 119.52&0.16&2.65  \\
       \midrule
    $x$-pred & $v$-loss &\cmark & 5.57 & 0.46 & 3.45\\
    $v$-pred & $v$-loss &\cmark& 3.93 & 0.48& 2.94 \\
    \bottomrule
  \end{tabular}
  } \vspace{-0.5cm}
\end{wraptable}
We investigate the prediction and loss spaces for E2E TTS. Unlike previous CFM-based waveform generation models, which generate waveforms conditioned on Mel-spectrograms, E2E TTS must directly learn the energy distribution of target waveforms without relying on ground-truth acoustic conditioning during generation.
Although $v$-prediction yields slightly better text-speech alignment and speaker similarity where $v$-prediction focuses low-frequency information due to noisy target velocity, we observe that it lacks the ability to accurately estimate waveform energy, resulting in significant degradation in perceptual quality. In contrast, Table \ref{table:loss} shows that $x$-prediction generates waveforms with more accurate energy distribution and significantly improves the perceptual quality of speech. This observation is particularly important because it suggests that high-quality waveform generation can be achieved without adversarial training. Note that the $x$-loss space introduces significant metallic noise and should therefore be avoided in current systems. 

\subsection{Multi-scale Spectral $v$-loss} \vspace{-0.2cm}
\begin{wraptable}{r}{7.5cm} \vspace{-0.7cm}
  \caption{Loss Comparison}
  \label{vloss}
  \centering
 \resizebox{7.5cm}{!}{
  \begin{tabular}{l|l|ccc}
    \toprule
 Method & Loss  &  WER & SIM & DNSMOS    \\
   \midrule
 Orig. CFM \cite{lee2025periodwave} &   $\mathcal{L}_{v}$ & 5.57 & 0.46 & 3.45 \\
 Modified CFM \cite{yao2026flowgan} &   $\mathcal{L}_{v\text{-}spec}$ & 68.82 & 0.46 & 3.49\\
 Ours (Ablation) & $\mathcal{L}_{ms\text{-}v\text{-}spec}$ & 16.98 & 0.51 & 3.57 \\
   \midrule
 Ours (Final) &   $\mathcal{L}_{v}$ + $\lambda_{ms}$$\mathcal{L}_{ms\text{-}v\text{-}spec}$ &  4.96 & 0.47 & 3.72 \\

    \bottomrule
  \end{tabular}
  } \vspace{-0.5cm}
\end{wraptable}

To further improve high-frequency detail reconstruction in CFM-based waveform generation, we investigate an enhanced CFM objective in the spectral domain. Flow2GAN \cite{yao2026flowgan} introduced a novel spectral loss that minimizes waveform errors with spectral energy scaling, enabling balanced frequency-wise optimization. This differs from conventional spectrogram losses, which directly minimize the discrepancy between predicted and target spectrograms. However, we found that such a loss tends to overemphasize high-frequency details at the expense of low-frequency structure, which can make text-speech alignment more difficult in E2E TTS. Unlike neural vocoders, E2E TTS models cannot rely on ground-truth spectral inputs during generation. To address this issue, we propose a multi-scale spectral $v$-loss that balances time-frequency optimization across multiple resolutions. The proposed loss improves overall frequency details and leads to better speaker similarity and perceptual quality as reported in Table \ref{vloss}. The final training objective combines the original CFM losses for wave and spec domains with our proposed multi-scale spectral $v$-loss.

\vspace{-0.2cm}
\subsection{Preliminary Study on REPA}\vspace{-0.2cm}
\begin{wraptable}{r}{7.5cm} \vspace{-0.65cm}
  \caption{REPA result}
  \label{table:repa}
  \centering
 \resizebox{7.5cm}{!}{
  \begin{tabular}{l|l|ccc}
    \toprule
 Method & Loss  &  WER & SIM & DNSMOS    \\
   \midrule
 Ours (Final) &   $\mathcal{L}_{v}$ + $\lambda_{ms}$$\mathcal{L}_{ms\text{-}v\text{-}spec}$ &  4.96 & 0.47 & 3.72 \\
   \midrule
  +REPA (Optional) & +$\lambda_{repa}$$\mathcal{L}_{repa}$ (Whisper) & 3.54 & 0.54 & 3.70 \\
    \bottomrule
  \end{tabular}
  } \vspace{-0.45cm}
\end{wraptable}
The results reported for our main WaveGen models do not use representation alignment (REPA) \cite{yu2024representation} or any external modules. However, recent studies on REPA suggest that external representation guidance can improve diffusion training. We conduct a small-scale preliminary experiment (WaveGen-S, LibriTTS) to examine whether REPA can further improve E2E waveform generation.
We apply REPA to the middle layer (10th layer) using semantic representations extracted from the encoder output of Whisper-small, and adopt a cosine similarity-based alignment loss weighted by $\lambda_{repa}=$ 0.1. As shown in Table \ref{table:repa}, REPA improves WER and SIM while maintaining comparable DNSMOS, suggesting faster convergence and improved training stability. We leave further improvements with REPA for future work.

\vspace{-0.2cm}
\subsection{Scalability}\vspace{-0.2cm}
\begin{table}[h]
  \caption{Model comparison with respect to model size and dataset scale} \vspace{-0.3cm}
  \label{table:scalability}
  \centering
\resizebox{1\textwidth}{!}{
  \begin{tabular}{l|ccc|ccc|ccc}
    \toprule
    Model & Params. & \makecell{Data \\ (hours)} &  \makecell{Training \\ Steps}   &  \makecell{\\ WER $\downarrow$} &   \makecell{Basic CFG (2.0)\\SIM $\uparrow$} &  \makecell{\\DNSMOS $\uparrow$} &  \makecell{\\ WER $\downarrow$} &   \makecell{Modified CFG (2.0)\\SIM $\uparrow$} &  \makecell{\\DNSMOS $\uparrow$}    \\
   \midrule
    WaveGen-S & 113M & 0.5k & 1.1M & 4.96 & 0.47 & 3.72 & 4.64 & 0.49& 3.57 \\
    WaveGen-B & 407M & 0.5k & 1.1M & 2.91 & 0.53& 3.67 & 2.93&0.55&3.59 \\
   \midrule
    WaveGen-S & 113M & 120k & 1.1M & 4.27 &  0.49 & 3.60 &4.42&0.53&3.56\\
    WaveGen-B & 407M & 120k & 1.1M & 2.37& 0.59& 3.60 & 1.77&0.63 & 3.60\\
    \bottomrule
  \end{tabular}
  } \vspace{-0.2cm}
\end{table}

We investigate the scalability of E2E TTS models with respect to both model size and dataset scale. Specifically, we train two model variants, S and B, with 0.1B and 0.4B parameters, respectively. Each model is trained on both LibriTTS (0.5k hours) and Emilia (120k hours). As shown in Table \ref{table:scalability}, the proposed models show consistent improvement in intelligibility and similarity as both model size and dataset scale increase. Furthermore, we find that the modified CFG strategy improves speaker similarity, suggesting that the initial sampling steps play an important role in adapting speaker characteristics. We provide additional results with different training steps in Table \ref{table:trainingsteps}. Based on the trends observed from extended training and the REPA results in Table \ref{table:repa}, we believe that further training can improve the overall quality of our models. However, due to limited computational resources, we train the models for only four epochs in this study.

\begin{wraptable}{r}{6.0cm} \vspace{-0.7cm}
  \caption{Analysis of NFE (Seed-en)}
  \label{table:NFE}
  \centering
 \resizebox{6.0cm}{!}{
  \begin{tabular}{l|c|ccc}
    \toprule
 Model & NFE  &  WER & SIM & DNSMOS    \\
   \midrule
 &  4   & 52.0 & 0.39 & 3.06 \\
 &  8   &  4.85 & 0.61 & 3.48 \\
 &  16   &  2.16 & 0.63 & 3.56 \\
 WaveGen-B &  32   &   1.77&0.63 & 3.60 \\
 &  50   & 1.87 & 0.64 & 3.61\\
 &  100   &  1.84 & 0.64 & 3.63 \\
    \bottomrule
  \end{tabular}
  } \vspace{-0.45cm}
\end{wraptable}

We further investigate the effect of NFE on E2E TTS. Table \ref{table:NFE} presents the results of WaveGen-B with different NFE settings. We find that speaking style is adapted in the early sampling stages, and high-quality waveform generation requires at least eight steps. Increasing the number of sampling steps generally improves similarity and perceptual quality. These results suggest that early sampling steps play a crucial role in establishing text-speech alignment and adapting speaking style, whereas additional steps mainly refine waveform.

\newpage
\begin{table}[t]
    \centering
    \caption{Multilingual TTS results using F5-TTS architecture and large-scale training configuration. For a fair comparison, all models are trained with the open-source Emilia dataset about 100k hours (English and Chinese). With end-to-end waveform generation, our models achieve the state-of-the-art performance on Seed-TTS-Eval benchmarks using public-available datasets.   } \vspace{-0.2cm}
    \label{tab:muliti_tts_seed_benchmark}
    \resizebox{1\linewidth}{!}{
    \begin{tabular}{lccc|clcc|ccc|ccc}
        \toprule 

        Model & E2E & SF (Drop) & Anchoring &  Hz & Hop (Frame) & Params & Data  & \makecell{ \\\textbf{WER} $\downarrow$} & \makecell{ \textbf{seed-en} \\ \textbf{SIM}$\uparrow$ } & \makecell{ \\\textbf{UTMOS} $\uparrow$}  & \makecell{ \\\textbf{CER} $\downarrow$} & \makecell{ \textbf{seed-zh} \\ \textbf{SIM}$\uparrow$ } & \makecell{ \\\textbf{UTMOS} $\uparrow$} \\
        \midrule
        GT& - & - & -  &  24k & - & - & -  & 2.06&0.73&3.53&1.26&0.76&2.78\\
        \midrule
        Seed-TTS \cite{anastassiou2024seed} & \xmark & \xmark & - & 24k & - & - & - & 2.25& 0.76&- & 1.12&0.79 & - \\
        VoxCPM-Emilia \cite{zhou2025voxcpm} & \xmark & \xmark & -& 16k & - & 0.5B & 100K & 2.34& 0.68 & - &1.11& 0.74 & - \\ 
        MaskGCT \cite{wang2025maskgct} & \xmark & \xmark & - & 24k & - & 2.2B & 100K & 2.88&0.71&3.55&2.40&0.77&2.63 \\
        OmniVoice-Emilia \cite{zhu2026omnivoice} & \xmark & \xmark & -& 24k & - & 0.8B & 100K & 1.72&0.71&3.88&0.89&0.76&3.05 \\
        \midrule
        ZipVoice \cite{zhu2025zipvoice} & \xmark & \xmark & - & 24k & 256 (93.75) & 123M & 100K& 1.70&0.69&3.82&1.40&0.75&3.15\\
        E2-TTS \cite{eskimez2024e2}  & \xmark & \xmark & - & 24k & 256 (93.75) & 333M & 100K   & 2.19 & 0.71& 3.20 &1.97& 0.73& 2.27 \\
        F5-TTS \cite{chen2024f5}  & \xmark & \xmark & -   & 24k & 256 (93.75) & 336M & 100K & 1.83 & 0.67 & 3.76 & 1.56 & 0.76&2.95\\
        \midrule
        BareWave \cite{fan2026barewave}  & \cmark & \xmark & - & 24k & 768 (31.25) & 983M & 19.4K  & 1.75 &0.60&3.72 & - & - & - \\
        WavTTS \cite{chen2026wavtts}  & \cmark & \xmark & - & 16k & 160 (100) & 673M & 100K  & 1.50&0.65&3.92&1.59&0.73&3.08 \\
        WaveGen-16k  & \cmark & \cmark (0.5) & \xmark & 16k &160 (100) & 676M & 100K  & 1.51 & 0.71 & 3.90 &1.65 & 0.77 &3.09 \\
        WaveGen-16k  & \cmark & \cmark (0.5) & \cmark & 16k &160 (100) & 676M & 100K  & 1.69 & 0.75 & 3.79 & 1.92 & 0.80 &3.01 \\
        \midrule
        WaveGen  & \cmark & \cmark (0.2) & \xmark & 24k & 480 (50) & 680M & 100K & 1.55 & 0.72 & 3.68 & 1.66& 0.78&2.84\\
        WaveGen  & \cmark & \cmark (0.2) & \cmark & 24k & 480 (50) & 680M & 100K & 1.47& 0.75& 3.68 & 1.75&0.80&2.87\\
        
        WaveGen (Ablation) & \cmark & \cmark (0) & \xmark & 24k & 480 (50) & 680M & 100K  & 2.07&0.73&3.60 & 2.33&0.79&2.76 \\
        WaveGen (Ablation)& \cmark & \cmark (0) & \cmark & 24k & 480 (50) & 680M & 100K  & 1.97&0.75&3.70 & 2.11&0.80&2.85 \\

        WaveGen (Ablation)& \cmark & \cmark (0.5) & \xmark & 24k & 480 (50) & 680M & 100K  & 1.39&0.70&3.81 & 1.58&0.76&2.97 \\
        WaveGen (Ablation)& \cmark & \cmark (0.5) & \cmark & 24k & 480 (50) & 680M & 100K & 1.56& 0.74 & 3.70 & 1.74&0.79&2.87 \\

        \bottomrule
    \end{tabular}
    } \vspace{-0.3cm}
\end{table}
\subsection{Multilingual TTS results } \vspace{-0.2cm}
We conduct large-scale multilingual TTS experiments on the English (En) and Chinese (Zh) subsets of the Emilia dataset, and adopt F5-TTS-based architectures including ConvNeXt-based text encoder and filler token-based conditioning and hyperparameters of WavTTS including waveform scaling and time scheduling for fair comparison.\footnote{To the best of our knowledge, WaveGen is the first CFM-based end-to-end waveform TTS model to outperform conventional multi-stage TTS systems on the Emilia dataset. WaveGen also achieves the best performance among the evaluated waveform diffusion TTS models, including WavTTS and BareWave, both of which were released after our NeurIPS 2026 submission. Nevertheless, we acknowledge that WavTTS provides valuable analyses of optimization strategies for waveform diffusion, which complement our findings and offer useful insights into improving the training of waveform TTS systems.} Table \ref{tab:muliti_tts_seed_benchmark} shows that WaveGen achieves strong performance across all evaluation metrics, demonstrating the effectiveness of direct waveform modeling and spectral forcing. Specifically, WaveGen achieves speaker similarities of 0.75 and 0.80 on the seed-en and seed-zh benchmark using only the publicly available Emilia dataset. Notably, we do not employ REPA when training WaveGen, suggesting further room for improvement through additional representation alignment techniques. In the following subsections and Appendix \ref{sec:multi-lingual-wavegen}, we further provide detailed analyses of the waveform diffusion optimization techniques used in WaveGen.

\vspace{-0.2cm}
\subsection{Information bottleneck pre-net for multi-representation generation}\vspace{-0.2cm}
We further investigate the effect of Spectral Forcing during large-scale training. When using a large-scale dataset, the dependency of Spectral Forcing increases because the model has difficulty aligning text with speech at the initial stage. While Spectral Forcing is weak guidance, the model is prone to learn to reconstruct waveforms from Spectral Forcing, resulting in delayed emergence of text-speech alignment. To reduce this limitation, we revisit the simple techniques, pre-net used in Tacotron \cite{wang17n_interspeech} by adding information bottleneck layers with two MLP layers and dropout to reduce dependency on teacher forcing for AR training. Specifically, we utilize 8n Mel bins and smaller bottleneck dimension of 32 similar to \cite{chen20r_interspeech}. We apply this bottleneck layer with dropout of 0.2 during training and inference. It improves the robustness of multi-representation generation by learning to generate waveform from multiple conditions. We investigated the effect of dropout ratio on Spectral Forcing in Appendix \ref{sec:spectral_dropout}. 

\vspace{-0.2cm}
\subsection{Prompt Anchor}\vspace{-0.2cm}
Following \cite{xin2026longcat}, we observe a train-inference mismatch in iterative generation under masked-infilling training. During training, the prompt region is fixed to the GT speech and is therefore not explicitly optimized for iterative updates. At inference time, however, repeatedly updating $x_t$ in the prompt region can introduce undesirable deviations from the GT prompt trajectory.

To address this issue, we anchor the prompt region to a GT-based intermediate state during iterative sampling. Specifically, at each sampling step, we construct the prompt-region $x_t$ by interpolating between the corresponding $x_0$ and $x_1$, rather than using the model-updated values. This simple prompt anchoring strategy substantially improves speaker similarity by reducing the train-inference mismatch introduced by masked-infilling training. 

\newpage
\begin{figure}[h]
    \centering
    {\includegraphics[width=1\textwidth]{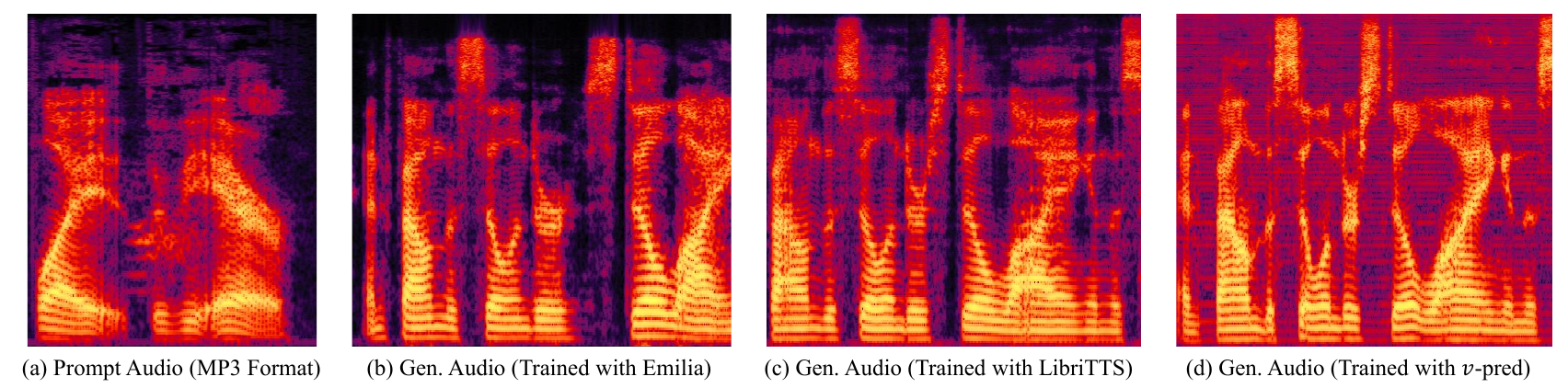}} \vspace{-0.7cm}
    \caption{Spectral analysis with respect to audio format and prediction space} \vspace{-0.7cm}
    \label{fig2:spec}
\end{figure}    

\section{Spectral Analysis of Audio Format and Prediction Space}\label{sec5} \vspace{-0.3cm}
\paragraph{Impact of Audio Compression}  

Neural vocoders and audio codecs are commonly trained on high-quality waveform datasets, and their training can often be effective even with relatively small datasets. By contrast, modern TTS models require large-scale training datasets to model diverse speaking styles including prosody and timbre in zero-shot scenarios. Although Emilia enables large-scale training, its MP3-compressed audio can limit the amount of recoverable high-frequency detail, which is particularly important for end-to-end waveform generation. As a result, models trained on such data tend to struggle with generating high-frequency components above 8 kHz. In contrast, models trained on LibriTTS preserve and generate richer high-frequency information as illustrated in Figure \ref{fig2:spec}.
This observation highlights the importance of carefully curating training data for E2E TTS, where the model must directly learn both linguistic alignment and high-fidelity waveform generation.

\paragraph{Effect of Prediction Space} \vspace{-0.2cm}
$x$-prediction plays a crucial role in high-quality waveform generation. As shown in Figure \ref{fig2:spec}, the model trained with $v$-prediction fails to recover a reliable energy distribution because the target velocity is highly noisy, resulting in substantially degraded waveform quality.

\vspace{-0.2cm}
\section{Potential Broader Impact} \vspace{-0.2cm} \label{sec:potential_Broader_Impact}
\paragraph{Societal Positive Impact} Our work proposes an E2E TTS model that directly generates high-quality waveform from text, thereby simplifying existing complex multi-stage TTS pipelines and enhancing training and inference efficiency. This efficient simplification, which enables speech generation with a single model without relying on external modules, enhances the accessibility of speech technologies and provides a foundation for broader adoption in both academic and industrial settings.  
Furthermore, in terms of application technology, this method can have a positive impact on diverse fields such as voice interfaces, education, and media content. In particular, the generation of clear and natural speech can improve the user experience and contribute to the advancement of speech language model.  

\vspace{-0.2cm}

\paragraph{Societal Negative Impact} Although the model and method proposed in this work do not directly raise ethical issues, there is a possibility of misuse when combined with voice cloning due to their ability to generate high-quality speech. For example, the model could be used to imitate a specific speaker’s voice or generate realistic fake speech to convey misleading information, which may lead to issues such as voice phishing or impersonation.
To mitigate these potential risks, it is necessary to pursue complementary research on fake audio detection as well as speech safety, including methods for identification and authenticity verification. As part of future work, we plan to further explore watermarking and authentication techniques to enable tracing the origin of generated audio.

\vspace{-0.2cm}

\paragraph{Limitation} 
We introduce an E2E TTS model that does not rely on any external modules, including SSL models. Nevertheless, SSL-based guidance, such as REPA, can be incorporated to accelerate training. Due to limited computational resources, we restrict large-scale training to four epochs. The observed scaling trend suggests that longer training may further improve performance, which we leave for future work. In addition, we plan to extend our model to 48 kHz waveform generation.

\vspace{-0.2cm}

\section{Conclusion} \vspace{-0.2cm}
We proposed WaveGen, a fully E2E diffusion-based text-to-waveform TTS system. Unlike previous TTS systems, WaveGen directly generates raw waveforms from text without separately trained intermediate representations. By introducing Spectral Forcing, disentangled DiT heads, and a multi-scale spectral $v$-loss, WaveGen enables stable CFM-based optimization in the high-resolution waveform domain. Our results show that direct text-to-waveform generation can achieve promising zero-shot TTS performance and benefit from model and dataset scaling, suggesting a simple and scalable alternative to increasingly complex cascaded TTS pipelines.
\newpage

\section{Acknowledgments}
This work was supported by Institute of Information \& communications Technology Planning \& Evaluation (IITP) grant funded by the Korea government (MSIT) (RS-2025-02283048, Developing the Next-Generation General AI with Reliability, Ethics, and Adaptability, IITP-2026-RS-2023-00255968, the Artificial Intelligence Convergence Innovation Human Resources Development, RS-2026-25617480,  AI Star Fellowship Support
Program(Gyeongsang National University)), National Research Foundation of Korea (NRF) grant funded by the Korea government (MSIT) (RS-2026-25518808), the “Advanced GPU Utilization Support Program” funded by the Government of the Republic of Korea (Ministry of Science and ICT), and Artificial intelligence industrial convergence cluster development project funded by the Ministry of Science and ICT (MSIT, Korea)\&Gwangju Metropolitan City.

\bibliographystyle{plainnat}
\bibliography{biblio}

@article{baade2026latent,
  title={Latent forcing: Reordering the diffusion trajectory for pixel-space image generation},
  author={Baade, Alan and Chan, Eric Ryan and Sargent, Kyle and Chen, Changan and Johnson, Justin and Adeli, Ehsan and Fei-Fei, Li},
  journal={arXiv preprint arXiv:2602.11401},
  year={2026}
}

@article{li2025back,
  title={Back to basics: Let denoising generative models denoise},
  author={Li, Tianhong and He, Kaiming},
  journal={arXiv preprint arXiv:2511.13720},
  year={2025}
}

@article{yu2024representation,
  title={Representation alignment for generation: Training diffusion transformers is easier than you think},
  author={Yu, Sihyun and Kwak, Sangkyung and Jang, Huiwon and Jeong, Jongheon and Huang, Jonathan and Shin, Jinwoo and Xie, Saining},
  journal={arXiv preprint arXiv:2410.06940},
  year={2024}
}

@article{du2024cosyvoice,
  title={Cosyvoice 2: Scalable streaming speech synthesis with large language models},
  author={Du, Zhihao and Wang, Yuxuan and Chen, Qian and Shi, Xian and Lv, Xiang and Zhao, Tianyu and Gao, Zhifu and Yang, Yexin and Gao, Changfeng and Wang, Hui and others},
  journal={arXiv preprint arXiv:2412.10117},
  year={2024}
}

@article{oord2016wavenet,
  title={Wavenet: A generative model for raw audio},
  author={Oord, Aaron van den and Dieleman, Sander and Zen, Heiga and Simonyan, Karen and Vinyals, Oriol and Graves, Alex and Kalchbrenner, Nal and Senior, Andrew and Kavukcuoglu, Koray},
  journal={arXiv preprint arXiv:1609.03499},
  year={2016}
}

@inproceedings{
ping2018clarinet,
title={ClariNet: Parallel Wave Generation in End-to-End Text-to-Speech},
author={Wei Ping and Kainan Peng and Jitong Chen},
booktitle={International Conference on Learning Representations},
year={2019},
url={https://openreview.net/forum?id=HklY120cYm},
}

@article{kumar2019melgan,
  title={Melgan: Generative adversarial networks for conditional waveform synthesis},
  author={Kumar, Kundan and Kumar, Rithesh and De Boissiere, Thibault and Gestin, Lucas and Teoh, Wei Zhen and Sotelo, Jose and De Brebisson, Alexandre and Bengio, Yoshua and Courville, Aaron C},
  journal={Advances in neural information processing systems},
  volume={32},
  year={2019}
}

@inproceedings{yamamoto2020parallel,
  title={Parallel WaveGAN: A fast waveform generation model based on generative adversarial networks with multi-resolution spectrogram},
  author={Yamamoto, Ryuichi and Song, Eunwoo and Kim, Jae-Min},
  booktitle={ICASSP 2020-2020 IEEE International Conference on Acoustics, Speech and Signal Processing (ICASSP)},
  pages={6199--6203},
  year={2020},
  organization={IEEE}
}

@article{kong2020hifi,
  title={Hifi-gan: Generative adversarial networks for efficient and high fidelity speech synthesis},
  author={Kong, Jungil and Kim, Jaehyeon and Bae, Jaekyoung},
  journal={Advances in neural information processing systems},
  volume={33},
  pages={17022--17033},
  year={2020}
}

@inproceedings{
lee2023bigvgan,
title={Big{VGAN}: A Universal Neural Vocoder with Large-Scale Training},
author={Lee, Sang-gil and Wei Ping and Boris Ginsburg and Bryan Catanzaro and Sungroh Yoon},
booktitle={The Eleventh International Conference on Learning Representations },
year={2023},
url={https://openreview.net/forum?id=iTtGCMDEzS_}
}

@inproceedings{
siuzdak2024vocos,
title={Vocos: Closing the gap between time-domain and Fourier-based neural vocoders for high-quality audio synthesis},
author={Hubert Siuzdak},
booktitle={The Twelfth International Conference on Learning Representations},
year={2024},
url={https://openreview.net/forum?id=vY9nzQmQBw}
}

@article{zeghidour2021soundstream,
  title={Soundstream: An end-to-end neural audio codec},
  author={Zeghidour, Neil and Luebs, Alejandro and Omran, Ahmed and Skoglund, Jan and Tagliasacchi, Marco},
  journal={IEEE/ACM Transactions on Audio, Speech, and Language Processing},
  volume={30},
  pages={495--507},
  year={2021},
  publisher={IEEE}
}

@article{dfossez2023high,
title={High Fidelity Neural Audio Compression},
author={Alexandre D{\'e}fossez and Jade Copet and Gabriel Synnaeve and Yossi Adi},
journal={Transactions on Machine Learning Research},
issn={2835-8856},
year={2023},
url={https://openreview.net/forum?id=ivCd8z8zR2},
note={Featured Certification, Reproducibility Certification}
}

@article{wang2023neural,
  title={Neural codec language models are zero-shot text to speech synthesizers},
  author={Wang, Chengyi and Chen, Sanyuan and Wu, Yu and Zhang, Ziqiang and Zhou, Long and Liu, Shujie and Chen, Zhuo and Liu, Yanqing and Wang, Huaming and Li, Jinyu and others},
  journal={arXiv preprint arXiv:2301.02111},
  year={2023}
}

@article{yang2023uniaudio,
  title={Uniaudio: An audio foundation model toward universal audio generation},
  author={Yang, Dongchao and Tian, Jinchuan and Tan, Xu and Huang, Rongjie and Liu, Songxiang and Chang, Xuankai and Shi, Jiatong and Zhao, Sheng and Bian, Jiang and Wu, Xixin and others},
  journal={arXiv preprint arXiv:2310.00704},
  year={2023}
}

@inproceedings{
roman2023from,
title={From Discrete Tokens to High-Fidelity Audio Using Multi-Band Diffusion},
author={Robin San Roman and Yossi Adi and Antoine Deleforge and Romain Serizel and Gabriel Synnaeve and Alexandre D{\'e}fossez},
booktitle={Thirty-seventh Conference on Neural Information Processing Systems},
year={2023},
url={https://openreview.net/forum?id=dOanKg3jKS}
}

@inproceedings{shen2018natural,
  title={Natural tts synthesis by conditioning wavenet on mel spectrogram predictions},
  author={Shen, Jonathan and Pang, Ruoming and Weiss, Ron J and Schuster, Mike and Jaitly, Navdeep and Yang, Zongheng and Chen, Zhifeng and Zhang, Yu and Wang, Yuxuan and Skerrv-Ryan, Rj and others},
  booktitle={2018 IEEE international conference on acoustics, speech and signal processing (ICASSP)},
  pages={4779--4783},
  year={2018},
  organization={IEEE}
}

@article{ren2019fastspeech,
  title={Fastspeech: Fast, robust and controllable text to speech},
  author={Ren, Yi and Ruan, Yangjun and Tan, Xu and Qin, Tao and Zhao, Sheng and Zhao, Zhou and Liu, Tie-Yan},
  journal={Advances in neural information processing systems},
  volume={32},
  year={2019}
}

@article{kim2020glow,
  title={Glow-tts: A generative flow for text-to-speech via monotonic alignment search},
  author={Kim, Jaehyeon and Kim, Sungwon and Kong, Jungil and Yoon, Sungroh},
  journal={Advances in Neural Information Processing Systems},
  volume={33},
  pages={8067--8077},
  year={2020}
}

@inproceedings{NEURIPS2023_2d8911db,
 author = {Le, Matthew and Vyas, Apoorv and Shi, Bowen and Karrer, Brian and Sari, Leda and Moritz, Rashel and Williamson, Mary and Manohar, Vimal and Adi, Yossi and Mahadeokar, Jay and Hsu, Wei-Ning},
 booktitle = {Advances in Neural Information Processing Systems},
 doi = {10.52202/075280-0618},
 editor = {A. Oh and T. Naumann and A. Globerson and K. Saenko and M. Hardt and S. Levine},
 pages = {14005--14034},
 publisher = {Curran Associates, Inc.},
 title = {Voicebox: Text-Guided Multilingual Universal Speech Generation at Scale},
 url = {https://proceedings.neurips.cc/paper_files/paper/2023/file/2d8911db9ecedf866015091b28946e15-Paper-Conference.pdf},
 volume = {36},
 year = {2023}
}

@inproceedings{NEURIPS2023_eb0965da,
 author = {Kim, Sungwon and Shih, Kevin and badlani, rohan and Santos, Joao Felipe and Bakhturina, Evelina and Desta, Mikyas and Valle, Rafael and Yoon, Sungroh and Catanzaro, Bryan},
 booktitle = {Advances in Neural Information Processing Systems},
 doi = {10.52202/075280-3246},
 editor = {A. Oh and T. Naumann and A. Globerson and K. Saenko and M. Hardt and S. Levine},
 pages = {74213--74228},
 publisher = {Curran Associates, Inc.},
 title = {P-Flow: A Fast and Data-Efficient Zero-Shot TTS through Speech Prompting},
 url = {https://proceedings.neurips.cc/paper_files/paper/2023/file/eb0965da1d2cb3fbbbb8dbbad5fa0bfc-Paper-Conference.pdf},
 volume = {36},
 year = {2023}
}

@inproceedings{
kreuk2023audiogen,
title={AudioGen: Textually Guided Audio Generation},
author={Felix Kreuk and Gabriel Synnaeve and Adam Polyak and Uriel Singer and Alexandre D{\'e}fossez and Jade Copet and Devi Parikh and Yaniv Taigman and Yossi Adi},
booktitle={The Eleventh International Conference on Learning Representations },
year={2023},
url={https://openreview.net/forum?id=CYK7RfcOzQ4}
}

@inproceedings{zen19_interspeech,
  author={Heiga Zen and Viet Dang and Rob Clark and Yu Zhang and Ron J. Weiss and Ye Jia and Zhifeng Chen and Yonghui Wu},
  title={{LibriTTS: A Corpus Derived from LibriSpeech for Text-to-Speech}},
  year=2019,
  booktitle={Proc. Interspeech 2019},
  pages={1526--1530},
  doi={10.21437/Interspeech.2019-2441},
  issn={2308-457X}
}

@inproceedings{kaneko2022istftnet,
  title={iSTFTNet: Fast and lightweight mel-spectrogram vocoder incorporating inverse short-time Fourier transform},
  author={Kaneko, Takuhiro and Tanaka, Kou and Kameoka, Hirokazu and Seki, Shogo},
  booktitle={ICASSP 2022-2022 IEEE International Conference on Acoustics, Speech and Signal Processing (ICASSP)},
  pages={6207--6211},
  year={2022},
  organization={IEEE}
}

@inproceedings{kim2021conditional,
  title={Conditional variational autoencoder with adversarial learning for end-to-end text-to-speech},
  author={Kim, Jaehyeon and Kong, Jungil and Son, Juhee},
  booktitle={International Conference on Machine Learning},
  pages={5530--5540},
  year={2021},
  organization={PMLR}
}

@article{lee2024ditto,
  title={DiTTo-TTS: Efficient and Scalable Zero-Shot Text-to-Speech with Diffusion Transformer},
  author={Lee, Keon and Kim, Dong Won and Kim, Jaehyeon and Cho, Jaewoong},
  journal={arXiv preprint arXiv:2406.11427},
  year={2024}
}

@article{lee2024accelerating,
  title={Accelerating High-Fidelity Waveform Generation via Adversarial Flow Matching Optimization},
  author={Lee, Sang-Hoon and Choi, Ha-Yeong and Lee, Seong-Whan},
  journal={arXiv preprint arXiv:2408.08019},
  year={2024}
}

@inproceedings{radford2023robust,
  title={Robust speech recognition via large-scale weak supervision},
  author={Radford, Alec and Kim, Jong Wook and Xu, Tao and Brockman, Greg and McLeavey, Christine and Sutskever, Ilya},
  booktitle={International conference on machine learning},
  pages={28492--28518},
  year={2023},
  organization={PMLR}
}

@inproceedings{
liu2025rfwave,
title={{RFW}ave: Multi-band Rectified Flow for Audio Waveform Reconstruction},
author={Peng Liu and Dongyang Dai and Zhiyong Wu},
booktitle={The Thirteenth International Conference on Learning Representations},
year={2025},
url={https://openreview.net/forum?id=gRmWtOnTLK}
}

@inproceedings{eskimez2024e2,
  title={E2 tts: Embarrassingly easy fully non-autoregressive zero-shot tts},
  author={Eskimez, Sefik Emre and Wang, Xiaofei and Thakker, Manthan and Li, Canrun and Tsai, Chung-Hsien and Xiao, Zhen and Yang, Hemin and Zhu, Zirun and Tang, Min and Tan, Xu and others},
  booktitle={2024 IEEE Spoken Language Technology Workshop (SLT)},
  pages={682--689},
  year={2024},
  organization={IEEE}
}

@inproceedings{
lee2025periodwave,
title={PeriodWave: Multi-Period Flow Matching for High-Fidelity Waveform Generation},
author={Sang-Hoon Lee and Ha-Yeong Choi and Seong-Whan Lee},
booktitle={The Thirteenth International Conference on Learning Representations},
year={2025},
url={https://openreview.net/forum?id=tQ1PmLfPBL}
}

@article{chen2024f5,
  title={F5-tts: A fairytaler that fakes fluent and faithful speech with flow matching},
  author={Chen, Yushen and Niu, Zhikang and Ma, Ziyang and Deng, Keqi and Wang, Chunhui and Zhao, Jian and Yu, Kai and Chen, Xie},
  journal={arXiv preprint arXiv:2410.06885},
  year={2024}
}

@article{defossez2024moshi,
  title={Moshi: a speech-text foundation model for real-time dialogue},
  author={D{\'e}fossez, Alexandre and Mazar{\'e}, Laurent and Orsini, Manu and Royer, Am{\'e}lie and P{\'e}rez, Patrick and J{\'e}gou, Herv{\'e} and Grave, Edouard and Zeghidour, Neil},
  journal={arXiv preprint arXiv:2410.00037},
  year={2024}
}

@article{hu2026qwen3,
  title={Qwen3-TTS Technical Report},
  author={Hu, Hangrui and Zhu, Xinfa and He, Ting and Guo, Dake and Zhang, Bin and Wang, Xiong and Guo, Zhifang and Jiang, Ziyue and Hao, Hongkun and Guo, Zishan and others},
  journal={arXiv preprint arXiv:2601.15621},
  year={2026}
}

@inproceedings{
ren2021fastspeech,
title={FastSpeech 2: Fast and High-Quality End-to-End Text to Speech},
author={Yi Ren and Chenxu Hu and Xu Tan and Tao Qin and Sheng Zhao and Zhou Zhao and Tie-Yan Liu},
booktitle={International Conference on Learning Representations},
year={2021},
url={https://openreview.net/forum?id=piLPYqxtWuA}
}

@inproceedings{
donahue2021endtoend,
title={End-to-end Adversarial Text-to-Speech},
author={Jeff Donahue and Sander Dieleman and Mikolaj Binkowski and Erich Elsen and Karen Simonyan},
booktitle={International Conference on Learning Representations},
year={2021},
url={https://openreview.net/forum?id=rsf1z-JSj87}
}

@inproceedings{gao2023e3,
  title={E3 tts: Easy end-to-end diffusion-based text to speech},
  author={Gao, Yuan and Morioka, Nobuyuki and Zhang, Yu and Chen, Nanxin},
  booktitle={2023 IEEE Automatic Speech Recognition and Understanding Workshop (ASRU)},
  pages={1--8},
  year={2023},
  organization={IEEE}
}

@inproceedings{
zhou2026hierarchical,
title={Hierarchical Semantic-Acoustic Modeling via Semi-Discrete Residual Representations for Expressive End-to-End Speech Synthesis},
author={Yixuan Zhou and Guoyang Zeng and Xin Liu and Xiang Li and Renjie Yu and Ziyang Wang and Runchuan Ye and Weiyue Sun and Jiancheng Gui and Kehan Li and Zhiyong Wu and Zhiyuan Liu},
booktitle={The Fourteenth International Conference on Learning Representations},
year={2026},
url={https://openreview.net/forum?id=h5KLpGoqzC}
}

@inproceedings{
peng2026vibevoice,
title={VibeVoice: Expressive Podcast Generation with Next-Token Diffusion},
author={Zhiliang Peng and Jianwei Yu and Wenhui Wang and Yaoyao Chang and Yutao Sun and Li Dong and Yi Zhu and Weijiang Xu and Hangbo Bao and Zehua Wang and Shaohan Huang and Yan Xia and Furu Wei},
booktitle={The Fourteenth International Conference on Learning Representations},
year={2026},
url={https://openreview.net/forum?id=FihSkzyxdv}
}

@inproceedings{wang17n_interspeech,
  title     = {{Tacotron: Towards End-to-End Speech Synthesis}},
  author    = {Yuxuan Wang and R.J. Skerry-Ryan and Daisy Stanton and Yonghui Wu and Ron J. Weiss and Navdeep Jaitly and Zongheng Yang and Ying Xiao and Zhifeng Chen and Samy Bengio and Quoc Le and Yannis Agiomyrgiannakis and Rob Clark and Rif A. Saurous},
  year      = {2017},
  booktitle = {{Interspeech 2017}},
  pages     = {4006--4010},
  doi       = {10.21437/Interspeech.2017-1452},
  issn      = {2958-1796},
}

@inproceedings{li2019neural,
  title={Neural speech synthesis with transformer network},
  author={Li, Naihan and Liu, Shujie and Liu, Yanqing and Zhao, Sheng and Liu, Ming},
  booktitle={Proceedings of the AAAI conference on artificial intelligence},
  volume={33},
  number={01},
  pages={6706--6713},
  year={2019}
}

@article{du2025cosyvoice,
  title={Cosyvoice 3: Towards in-the-wild speech generation via scaling-up and post-training},
  author={Du, Zhihao and Gao, Changfeng and Wang, Yuxuan and Yu, Fan and Zhao, Tianyu and Wang, Hao and Lv, Xiang and Wang, Hui and Ni, Chongjia and Shi, Xian and others},
  journal={arXiv preprint arXiv:2505.17589},
  year={2025}
}

@inproceedings{popov2021grad,
  title={Grad-tts: A diffusion probabilistic model for text-to-speech},
  author={Popov, Vadim and Vovk, Ivan and Gogoryan, Vladimir and Sadekova, Tasnima and Kudinov, Mikhail},
  booktitle={International conference on machine learning},
  pages={8599--8608},
  year={2021},
  organization={PMLR}
}

@article{zhu2025zipvoice,
  title={Zipvoice: Fast and high-quality zero-shot text-to-speech with flow matching},
  author={Zhu, Han and Kang, Wei and Yao, Zengwei and Guo, Liyong and Kuang, Fangjun and Li, Zhaoqing and Zhuang, Weiji and Lin, Long and Povey, Daniel},
  journal={arXiv preprint arXiv:2506.13053},
  year={2025}
}

@article{gong2026moss,
  title={MOSS-TTS Technical Report},
  author={Gong, Yitian and Jiang, Botian and Zhao, Yiwei and Yuan, Yucheng and Chen, Kuangwei and Jiang, Yaozhou and Chang, Cheng and Hong, Dong and Chen, Mingshu and Li, Ruixiao and others},
  journal={arXiv preprint arXiv:2603.18090},
  year={2026}
}

@article{griffin1984signal,
  title={Signal estimation from modified short-time Fourier transform},
  author={Griffin, Daniel and Lim, Jae},
  journal={IEEE Transactions on acoustics, speech, and signal processing},
  volume={32},
  number={2},
  pages={236--243},
  year={1984},
  publisher={IEEE}
}

@inproceedings{
yao2026flowgan,
title={Flow2{GAN}: Hybrid Flow Matching and {GAN} with Multi-Resolution Network for Few-step High-Fidelity Audio Generation},
author={Zengwei Yao and Wei Kang and Han Zhu and Liyong Guo and Lingxuan Ye and Fangjun Kuang and Weiji Zhuang and Zhaoqing Li and Zhifeng Han and Long Lin and Daniel Povey},
booktitle={The Fourteenth International Conference on Learning Representations},
year={2026},
url={https://openreview.net/forum?id=5eTpRIULtb}
}

@inproceedings{NEURIPS2025_07134952,
 author = {Choi, Ha-Yeong and Lee, Sang-Hoon},
 booktitle = {Advances in Neural Information Processing Systems},
 doi = {10.52202/085713-0165},
 editor = {D. Belgrave and C. Zhang and H. Lin and R. Pascanu and P. Koniusz and M. Ghassemi and N. Chen},
 pages = {4669--4697},
 publisher = {Curran Associates, Inc.},
 title = {StreamFlow: Streaming Audio Generation from Discrete Tokens via Streaming Flow Matching},
 url = {https://proceedings.neurips.cc/paper_files/paper/2025/file/0713495297dab18fabca4795cb9ef8ef-Paper-Conference.pdf},
 volume = {38, Main Conference},
 year = {2025}
}

@inproceedings{peebles2023scalable,
  title={Scalable diffusion models with transformers},
  author={Peebles, William and Xie, Saining},
  booktitle={Proceedings of the IEEE/CVF international conference on computer vision},
  pages={4195--4205},
  year={2023}
}

@inproceedings{hai25_interspeech,
  title     = {{EzAudio: Enhancing Text-to-Audio Generation with Efficient Diffusion Transformer}},
  author    = {Jiarui Hai and Yong Xu and Hao Zhang and Chenxing Li and Helin Wang and Mounya Elhilali and Dong Yu},
  year      = {2025},
  booktitle = {{Interspeech 2025}},
  pages     = {4233--4237},
  doi       = {10.21437/Interspeech.2025-1137},
  issn      = {2958-1796},
}

@article{zhu2025zipvoicedia,
  title={Zipvoice-dialog: Non-autoregressive spoken dialogue generation with flow matching},
  author={Zhu, Han and Kang, Wei and Guo, Liyong and Yao, Zengwei and Kuang, Fangjun and Zhuang, Weiji and Li, Zhaoqing and Han, Zhifeng and Zhang, Dong and Zhang, Xin and others},
  journal={arXiv preprint arXiv:2507.09318},
  year={2025}
}

@article{zhu2026omnivoice,
  title={OmniVoice: Towards Omnilingual Zero-Shot Text-to-Speech with Diffusion Language Models},
  author={Zhu, Han and Ye, Lingxuan and Kang, Wei and Yao, Zengwei and Guo, Liyong and Kuang, Fangjun and Han, Zhifeng and Zhuang, Weiji and Lin, Long and Povey, Daniel},
  journal={arXiv preprint arXiv:2604.00688},
  year={2026}
}

@article{he2025emilia,
  title={Emilia: A large-scale, extensive, multilingual, and diverse dataset for speech generation},
  author={He, Haorui and Shang, Zengqiang and Wang, Chaoren and Li, Xuyuan and Gu, Yicheng and Hua, Hua and Liu, Liwei and Yang, Chen and Li, Jiaqi and Shi, Peiyang and others},
  journal={IEEE Transactions on Audio, Speech and Language Processing},
  year={2025},
  publisher={IEEE}
}

@inproceedings{
loshchilov2018decoupled,
title={Decoupled Weight Decay Regularization},
author={Ilya Loshchilov and Frank Hutter},
booktitle={International Conference on Learning Representations},
year={2019},
url={https://openreview.net/forum?id=Bkg6RiCqY7},
}

@article{wang2025spark,
  title={Spark-tts: An efficient llm-based text-to-speech model with single-stream decoupled speech tokens},
  author={Wang, Xinsheng and Jiang, Mingqi and Ma, Ziyang and Zhang, Ziyu and Liu, Songxiang and Li, Linqin and Liang, Zheng and Zheng, Qixi and Wang, Rui and Feng, Xiaoqin and others},
  journal={arXiv preprint arXiv:2503.01710},
  year={2025}
}

@article{peng2025vibevoice,
  title={Vibevoice technical report},
  author={Peng, Zhiliang and Yu, Jianwei and Wang, Wenhui and Chang, Yaoyao and Sun, Yutao and Dong, Li and Zhu, Yi and Xu, Weijiang and Bao, Hangbo and Wang, Zehua and others},
  journal={arXiv preprint arXiv:2508.19205},
  year={2025}
}

@article{zhou2025voxcpm,
  title={Voxcpm: Tokenizer-free TTS for context-aware speech generation and true-to-life voice cloning},
  author={Zhou, Yixuan and Zeng, Guoyang and Liu, Xin and Li, Xiang and Yu, Renjie and Wang, Ziyang and Ye, Runchuan and Sun, Weiyue and Gui, Jiancheng and Li, Kehan and others},
  journal={arXiv preprint arXiv:2509.24650},
  year={2025}
}

@inproceedings{NEURIPS2025_d8a12fde,
 author = {Wang, Yuancheng and Chen, Dekun and Zhang, Xueyao and Zhang, Junan and Li, Jiaqi and Wu, Zhizheng},
 booktitle = {Advances in Neural Information Processing Systems},
 doi = {10.52202/085713-4932},
 editor = {D. Belgrave and C. Zhang and H. Lin and R. Pascanu and P. Koniusz and M. Ghassemi and N. Chen},
 pages = {147494--147523},
 publisher = {Curran Associates, Inc.},
 title = {TaDiCodec: Text-aware Diffusion Speech Tokenizer for Speech Language Modeling},
 url = {https://proceedings.neurips.cc/paper_files/paper/2025/file/d8a12fde9e72444e1b356e8c37e53753-Paper-Conference.pdf},
 volume = {38, Main Conference},
 year = {2025}
}

@inproceedings{NEURIPS2025_be8987a7,
 author = {Wang, Yuancheng and Zheng, Jiachen and Zhang, Junan and Zhang, Xueyao and Liao, Huan and Wu, Zhizheng},
 booktitle = {Advances in Neural Information Processing Systems},
 doi = {10.52202/085713-4381},
 editor = {D. Belgrave and C. Zhang and H. Lin and R. Pascanu and P. Koniusz and M. Ghassemi and N. Chen},
 pages = {131615--131645},
 publisher = {Curran Associates, Inc.},
 title = {Metis: A Foundation Speech Generation Model with Masked Generative Pre-training},
 url = {https://proceedings.neurips.cc/paper_files/paper/2025/file/be8987a75afeeadfa65007cf7ee25de0-Paper-Conference.pdf},
 volume = {38, Main Conference},
 year = {2025}
}

@inproceedings{reddy2021dnsmos,
  title={DNSMOS: A non-intrusive perceptual objective speech quality metric to evaluate noise suppressors},
  author={Reddy, Chandan KA and Gopal, Vishak and Cutler, Ross},
  booktitle={ICASSP 2021-2021 IEEE International Conference on Acoustics, Speech and Signal Processing (ICASSP)},
  pages={6493--6497},
  year={2021},
  organization={IEEE}
}

@inproceedings{
welker2025flowdec,
title={FlowDec: A flow-based full-band general audio codec with high perceptual quality},
author={Simon Welker and Matthew Le and Ricky T. Q. Chen and Wei-Ning Hsu and Timo Gerkmann and Alexander Richard and YI-CHIAO WU},
booktitle={The Thirteenth International Conference on Learning Representations},
year={2025},
url={https://openreview.net/forum?id=uxDFlPGRLX}
}

@article{kumar2023high,
  title={High-fidelity audio compression with improved rvqgan},
  author={Kumar, Rithesh and Seetharaman, Prem and Luebs, Alejandro and Kumar, Ishaan and Kumar, Kundan},
  journal={Advances in Neural Information Processing Systems},
  volume={36},
  pages={27980--27993},
  year={2023}
}

@article{copet2023simple,
  title={Simple and controllable music generation},
  author={Copet, Jade and Kreuk, Felix and Gat, Itai and Remez, Tal and Kant, David and Synnaeve, Gabriel and Adi, Yossi and D{\'e}fossez, Alexandre},
  journal={Advances in neural information processing systems},
  volume={36},
  pages={47704--47720},
  year={2023}
}

@article{anastassiou2024seed,
  title={Seed-tts: A family of high-quality versatile speech generation models},
  author={Anastassiou, Philip and Chen, Jiawei and Chen, Jitong and Chen, Yuanzhe and Chen, Zhuo and Chen, Ziyi and Cong, Jian and Deng, Lelai and Ding, Chuang and Gao, Lu and others},
  journal={arXiv preprint arXiv:2406.02430},
  year={2024}
}

@inproceedings{
ji2025wavtokenizer,
title={WavTokenizer: an Efficient Acoustic Discrete Codec Tokenizer for Audio Language Modeling},
author={Shengpeng Ji and Ziyue Jiang and Wen Wang and Yifu Chen and Minghui Fang and Jialong Zuo and Qian Yang and Xize Cheng and Zehan Wang and Ruiqi Li and Ziang Zhang and Xiaoda Yang and Rongjie Huang and Yidi Jiang and Qian Chen and Siqi Zheng and Zhou Zhao},
booktitle={The Thirteenth International Conference on Learning Representations},
year={2025},
url={https://openreview.net/forum?id=yBlVlS2Fd9}
}

@inproceedings{
tian2026ualm,
title={{UALM}: Unified Audio Language Model for Understanding, Generation and Reasoning},
author={Jinchuan Tian and Sang-gil Lee and Zhifeng Kong and Sreyan Ghosh and Arushi Goel and Chao-Han Huck Yang and Wenliang Dai and Zihan Liu and Hanrong Ye and Shinji Watanabe and Mohammad Shoeybi and Bryan Catanzaro and Rafael Valle and Wei Ping},
booktitle={The Fourteenth International Conference on Learning Representations},
year={2026},
url={https://openreview.net/forum?id=TsdlOjcQNu}
}

@inproceedings{ye2025codec,
  title={Codec does matter: Exploring the semantic shortcoming of codec for audio language model},
  author={Ye, Zhen and Sun, Peiwen and Lei, Jiahe and Lin, Hongzhan and Tan, Xu and Dai, Zheqi and Kong, Qiuqiang and Chen, Jianyi and Pan, Jiahao and Liu, Qifeng and others},
  booktitle={Proceedings of the AAAI Conference on Artificial Intelligence},
  volume={39},
  number={24},
  pages={25697--25705},
  year={2025}
}

@inproceedings{
kim2025efficient,
title={Efficient Generative Modeling with Residual Vector Quantization-Based Tokens},
author={Jaehyeon Kim and Taehong Moon and Keon Lee and Jaewoong Cho},
booktitle={Forty-second International Conference on Machine Learning},
year={2025},
url={https://openreview.net/forum?id=f8uPzuMNDq}
}

@inproceedings{
wang2026scaling,
title={Scaling Speech Tokenizers with Diffusion Autoencoders},
author={Yuancheng Wang and Zhenyu Tang and Yun Wang and Arthur Hinsvark and Yingru Liu and Yinghao Aaron Li and Kainan Peng and Junyi Ao and Mingbo Ma and Mike Seltzer and Qing He and Xubo Liu},
booktitle={The Fourteenth International Conference on Learning Representations},
year={2026},
url={https://openreview.net/forum?id=llMfmDtWka}
}

@inproceedings{NEURIPS2025_220e6dbe,
 author = {Della Libera, Luca and Paissan, Francesco and Subakan, Cem and Ravanelli, Mirco},
 booktitle = {Advances in Neural Information Processing Systems},
 doi = {10.52202/085713-0801},
 editor = {D. Belgrave and C. Zhang and H. Lin and R. Pascanu and P. Koniusz and M. Ghassemi and N. Chen},
 pages = {23742--23767},
 publisher = {Curran Associates, Inc.},
 title = {FocalCodec: Low-Bitrate Speech Coding via Focal Modulation Networks},
 url = {https://proceedings.neurips.cc/paper_files/paper/2025/file/220e6dbed88faa83bd9f9e8fa359e43d-Paper-Conference.pdf},
 volume = {38, Main Conference},
 year = {2025}
}

@article{della2026beyond,
  title={Beyond Fixed Frames: Dynamic Character-Aligned Speech Tokenization},
  author={Della Libera, Luca and Subakan, Cem and Ravanelli, Mirco},
  journal={arXiv preprint arXiv:2601.23174},
  year={2026}
}

@inproceedings{chen20r_interspeech,
  title     = {{MultiSpeech: Multi-Speaker Text to Speech with Transformer}},
  author    = {Mingjian Chen and Xu Tan and Yi Ren and Jin Xu and Hao Sun and Sheng Zhao and Tao Qin},
  year      = {2020},
  booktitle = {{Interspeech 2020}},
  pages     = {4024--4028},
  doi       = {10.21437/Interspeech.2020-3139},
  issn      = {2958-1796},
}

@article{xin2026longcat,
  title={Longcat-audiodit: High-fidelity diffusion text-to-speech in the waveform latent space},
  author={Xin, Detai and Hu, Shujie and Yang, Chengzuo and Huang, Chen and Yu, Guoqiao and Wan, Guanglu and Cai, Xunliang},
  journal={arXiv preprint arXiv:2603.29339},
  year={2026}
}

@article{chen2026wavtts,
  title={WavTTS: Towards High-Quality Zero-Shot TTS via Direct Raw Waveform Modeling},
  author={Chen, Wenxi and Jia, Dongya and Chen, Yushen and Niu, Zhikang and Liang, Yuzhe and Li, Xiquan and Yan, Ruiqi and Ma, Ziyang and Yang, Guanrou and Chen, Sanyuan and others},
  journal={arXiv preprint arXiv:2606.03455},
  year={2026}
}

@article{fan2026barewave,
  title={BareWave: Waveform-Native Flow-Matching Text-to-Speech},
  author={Fan, Wei and Tan, Chao-Hong and Chen, Qian and Wang, Wen and Li, Xiangang and Chen, Kejiang and Zhang, Weiming and Yu, Nenghai},
  journal={arXiv preprint arXiv:2606.09048},
  year={2026}
}

@inproceedings{wang2025maskgct,
  title={Maskgct: Zero-shot text-to-speech with masked generative codec transformer},
  author={Wang, Yuancheng and Zhan, Haoyue and Liu, Liwei and Zeng, Ruihong and Guo, Haotian and Zheng, Jiachen and Zhang, Qiang and Zhang, Xueyao and Zhang, Shunsi and Wu, Zhizheng},
  booktitle={International Conference on Learning Representations},
  volume={2025},
  pages={47127--47150},
  year={2025}
}

@inproceedings{
lee2026regen,
title={ReGen: Hierarchical Multi-Prompt Representation Generation for Efficient Waveform Diffusion Models},
author={Sang-Hoon Lee and Ha-Yeong Choi},
booktitle={Forty-third International Conference on Machine Learning},
year={2026},
url={https://openreview.net/forum?id=T8Y9elIrXa}
}

@inproceedings{lee-choi-2026-hierarchical,
    title = "Hierarchical Representation Alignment Learning of Diffusion Transformers for Neural Audio Codec",
    author = "Lee, Sang-Hoon  and
      Choi, Ha-Yeong",
    editor = "Liakata, Maria  and
      Moreira, Viviane P.  and
      Zhang, Jiajun  and
      Jurgens, David",
    booktitle = "Findings of the {A}ssociation for {C}omputational {L}inguistics: {ACL} 2026",
    month = jul,
    year = "2026",
    address = "San Diego, California, United States",
    publisher = "Association for Computational Linguistics",
    url = "https://aclanthology.org/2026.findings-acl.1622/",
    doi = "10.18653/v1/2026.findings-acl.1622",
    pages = "32410--32426",
    ISBN = "979-8-89176-395-1"
}

\newpage
\appendix 

\section{Implementation Details}
\begin{table*}[htb!]
  \caption{Hyperparameters of WaveGen}
  \label{hyper}
  \centering
      \resizebox{1\textwidth}{!}{
  \begin{tabular}{c|l|ccc}
    \toprule
Module    &Hyperparameter   & WaveGen-S  &WaveGen-B  &WaveGen  \\
\midrule
 Waveform  & Sampling Rate &24000& 24000& 24000 \\
   & Reshape &(100, 240)& (100, 240) & (50,480) \\
   & Diffusion Resolution & 100Hz & 100Hz & 50Hz \\
\midrule
 Spectral Forcing & Hop &240& 240 & 480\\
                  & Window &960 &960 & 1920 \\
                  & FFT &960& 960 & 1920 \\
                  & Filter &Mel& Mel & Mel  \\
                  & Bin & 64& 64 & 80  \\
                  & IB & - & - & 32  \\
\midrule
 Text Embedding  & Dimension & 384 &384 & 512  \\
 & ConvNeXt & - & - & 4 \\
 & Conditioning & Average  & Average & Filler Token \\
\midrule
DiT  & Input Dim. & 768 & 1024 & 1158   \\
 & Hidden Dim. & 1536 & 4096 & 4632   \\
 & Shared Layer &  14 & 24 & 24  \\
 & Disentangled Layer & 2 & 2 & 2  \\
 & Head & 12 & 16 & 16   \\
 & Type & Scale-DiT & Scale-DiT & DiT \\
 & AdaLN &AdaLN-SOLA (rank=32) & AdaLN-SOLA (rank=32) & AdaLN-Zero  \\

 \midrule
Training&Training Step & 1.1M& 1.1M & 1.3M  \\
&Epoch&4&4&11\\
&GPU& 4 & 4 & 8 \\
&Learning Rate& $1\times10^{-4}$&$1\times10^{-4}$&$7.5\times10^{-5}$   \\
&Warm-up Scheduling & 20,000 steps & 20,000 steps & 20,000 steps  \\
&Learning Scheduling & Cosine Decay  & Cosine Decay & Linear Decay   \\
&$\lambda_{\text{sf}}$ &0.1 & 0.1 & 0.1 \\
&$\lambda_{\text{v-stft}}$ &  0.1& 0.1 & -  \\
&$\lambda_{\text{x-stft}}$ &  -& - & 0.05  \\
&Audio/Token Drop & 0.3/0.2 &0.3/0.2 & 0.1 (Joint) \\
 & Model Size& 113M & 407M & 680M   \\
 \midrule
Dataset&Max Frame & 480s &480s & 384s   \\
&Length & 3-30s& 3-30s & 3-30s \\
\midrule
t scheduling& - & Uniform  & Uniform  & Logit normal ($\mu$=-0.8, $\sigma$=0.8) \\
\midrule
Data Scaling& Waveform & Max Normalization + x0.95  & Max Normalization + x0.95  & x9 \\
\midrule
Noise Scaling& Waveform Noise &0.25  & 0.25 & 1 \\
&Spectral Noise &1 & 1 & 1 \\
\midrule
Sampling&ODE Method & Euler& Euler & Euler \\
&Steps & 32 & 32 & 50\\
&CFG & 2& 2 & 3 \\
\midrule
Objective&Prediction Space &  $x$-pred& $x$-pred & $x$-pred\\
&Loss Space & $v$-loss& $v$-loss & $v$-loss\\
&x to v conversion & $(x_\theta-x_t)/(1-t)$& $(x_\theta-x_t)/(1-t)$& $(x_\theta-x_t)/(1-t)$\\
& $t_{clip}$ & 0.05 & 0.05 & 0.02 \\
\midrule
Multi-scale  & Hop &[480, 240, 120, 60, 30, 15]& [480, 240, 120, 60, 30, 15] & -\\
 Spectral $v$-loss & Window &[1920, 960, 480, 240, 120, 60]& [1920, 960, 480, 240, 120, 60] & - \\
                  & FFT &[1920, 960, 480, 240, 120, 60] & [1920, 960, 480, 240, 120, 60] & - \\
                  & Filter &Linear & Linear & - \\
                  & Bin & [512,256,128,64,32,16] & [512,256,128,64,32,16] & - \\
\midrule
Multi-scale  & Hop & - & - & [512, 256, 128, 64, 32, 16, 8]  \\
  Spectral $x$-loss            & Window &- & - & [2048, 1024, 512, 256, 128, 64, 32] \\
                  & FFT &-& - & [2048, 1024, 512, 256, 128, 64, 32] \\
                  & Filter &- & - & Mel\\
                  & Bin & - & - & [320,160,80,40,20,10,5] \\
    \bottomrule
  \end{tabular}}
\end{table*}

\newpage
\section{Implementation Details for Baselines}
 
\paragraph{Spark-TTS}
Spark-TTS\footnote{\url{https://github.com/SparkAudio/Spark-TTS}} \cite{wang2025spark} is an LLM-based zero-shot TTS with BiCodec, a single-stream speech codec that represents speech with semantic tokens for linguistic content and global tokens for speaker attributes. 

\paragraph{TadiCodec-AR}  
TaDiCodec\footnote{\url{https://github.com/AmphionTeam/TaDiCodec}} \cite{NEURIPS2025_d8a12fde} is a text-aware diffusion speech tokenizer that compresses speech into a low frame rate discrete representation using a diffusion autoencoder with text-guided decoding. 

\paragraph{F5-TTS}
F5-TTS \cite{chen2024f5} is a non-autoregressive zero-shot TTS model based on flow matching with a Diffusion Transformer. It removes explicit duration modeling and phoneme-level alignment by padding text tokens to the speech length, and uses ConvNeXt-based text refinement to improve alignment robustness. For evaluation, we utilize the official implementation and released checkpoint.\footnote{\url{https://github.com/SWivid/F5-TTS}}

\paragraph{ZipVoice}
ZipVoice\footnote{\url{https://github.com/k2-fsa/ZipVoice}}  \cite{zhu2025zipvoice} is a compact zero-shot TTS model based on flow matching. It employs a Zipformer-based text encoder and vector-field estimator, uses average upsampling for initial speech-text alignment, and introduces flow distillation to reduce the number of sampling steps. 

\paragraph{CosyVoice2}
CosyVoice2\footnote{\url{https://github.com/FunAudioLLM/CosyVoice}}   \cite{du2024cosyvoice} is an LLM-based TTS model with a chunk-aware causal flow matching decoder. It improves speech token utilization with FSQ and supports both streaming and non-streaming synthesis within a unified framework.
 
\paragraph{CosyVoice3}
CosyVoice3 \cite{du2025cosyvoice} extends CosyVoice2 toward in-the-wild multilingual zero-shot TTS through larger-scale pre-training, a supervised multi-task speech tokenizer, and RL post-training. In our comparison, CosyVoice3-RL refers to the reward-optimized with differentiable reward optimization. For evaluation, we utilize the official implementation and released checkpoint.\footnote{\url{https://github.com/FunAudioLLM/CosyVoice}}

\paragraph{Qwen3-TTS}  
Qwen3-TTS \cite{hu2026qwen3} is an LLM-based TTS system built on the Qwen3 backbone. It employs a 12Hz speech tokenizer with a multi-codebook RVQ consisting of 16 codebooks to represent both semantic and acoustic information. To efficiently model the multi-codebook sequence, it adopts multi-token prediction decoding.
For evaluation, we utilize the official implementation and evaluate both the 0.6B and 1.7B checkpoints.\footnote{\url{https://github.com/QwenLM/Qwen3-TTS}}

\paragraph{MOSS-TTS}  
MOSS-TTS\footnote{\url{https://github.com/OpenMOSS/MOSS-TTS}} \cite{gong2026moss} employs the MOSS-Audio-Tokenizer to convert waveform into a unified sequence of discrete representations capturing both semantic and acoustic information.

\paragraph{VibeVoice}
VibeVoice \cite{peng2025vibevoice} is a long-form and multi-speaker TTS model designed to generate expressive conversational speech from text. It employs continuous speech tokenizers operating at low frame rate and uses next-token diffusion to autoregressively generate continuous speech representations. They release their open-source implementation\footnote{\url{https://github.com/microsoft/VibeVoice}} and real-time model\footnote{\url{https://huggingface.co/microsoft/VibeVoice-Realtime-0.5B}}.

\paragraph{VoxCPM}
VoxCPM\footnote{\url{https://github.com/OpenBMB/VoxCPM}} \cite{zhou2026hierarchical} uses hierarchical semantic-acoustic modeling, where semantic-prosodic planning and residual acoustic modeling are coupled with a local diffusion decoder for high-fidelity speech generation.

\newpage

\section{Evaluation Details} \vspace{-0.2cm}
\paragraph{WER} We calculate word error rate (WER) using open-source automatic speech recognition model, Whisper \cite{radford2023robust}. We utilize the official implementation of Whisper-large-v3\footnote{\url{https://github.com/openai/whisper}} to evaluate speech intelligibility.
\vspace{-0.2cm}
\paragraph{SIM} For evaluating speaker similarity quality, we utilized WavLM-large which is fine-tuned by speaker verification task\footnote{\url{https://drive.google.com/file/d/1-aE1NfzpRCLxA4GUxX9ITI3F9LlbtEGP/view}.}. We calculate cosine similarity between prompt and generated samples.
\vspace{-0.2cm}
\paragraph{DNSMOS} To assess the naturalness of the synthetic samples, we utilized the open-source Mean Opinion Score (MOS) prediction model, DNSMOS\footnote{\url{https://github.com/Lightning-AI/torchmetrics}}\cite{reddy2021dnsmos}. 

\section{Crowdsourcing Details} \label{appendix:mturk} \vspace{-0.2cm}
\begin{figure}
    \centering
    {\includegraphics[width=1\textwidth]{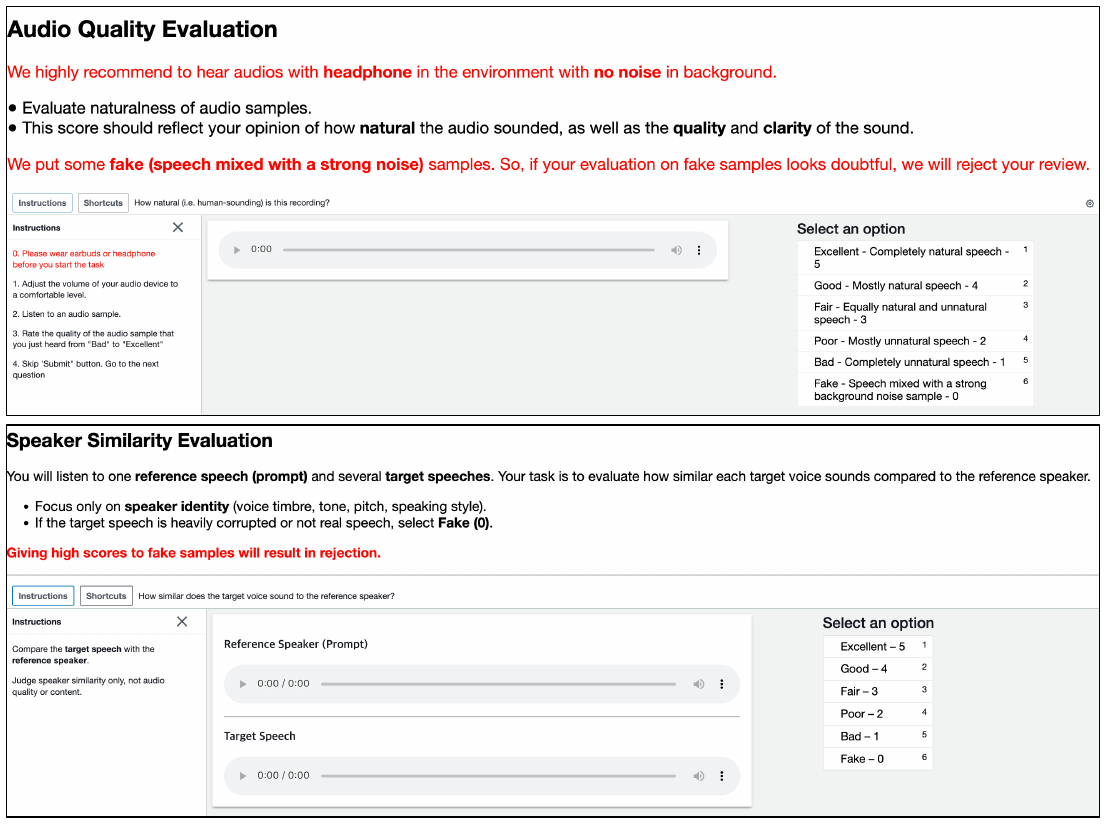}} 
    \caption{MOS interfaces}
    \label{fig:mturk} \vspace{-0.3cm}
\end{figure} 

We conducted Mean Opinion Score (MOS) evaluations using Amazon Mechanical Turk (MTurk)\footnote{\url{https://www.mturk.com/}} to qualitatively assess the quality of synthesized speech from the TTS. Specifically, we carried out two types of evaluations: nMOS, which measures the naturalness of the generated speech, and sMOS, which measures the similarity of the generated speaker timbre to that of the prompt speaker.
For each evaluation, 20 native speakers of American English participated. Each participant rated 100 samples per model on a 5-point scale ranging from 1 to 5. To reduce unreliable responses, such as participants randomly selecting scores, and to improve the reliability of the MTurk evaluation, we included Gaussian-noise samples as control samples. Responses from participants who failed to identify these samples as fake were excluded from the final analysis. In addition, only MTurk workers with a prior task approval rate of at least 80\% and at least 100 approved HITs were allowed to participate in the evaluation.

\newpage
\begin{table}
  \caption{Comparison with respect to training steps.}
  \label{table:trainingsteps}
  \centering
\resizebox{1\textwidth}{!}{
  \begin{tabular}{l|ccc|ccc|ccc}
    \toprule
    Model & Params. & Data &  \makecell{Training \\ Steps}   &  \makecell{\\ WER($\downarrow$)} &   \makecell{Basic CFG (2.0)\\SIM($\uparrow$)} &  \makecell{\\DNSMOS($\uparrow$)} &  \makecell{\\ WER($\downarrow$)} &   \makecell{No CFG\\SIM($\uparrow$)} &  \makecell{\\DNSMOS($\uparrow$)}    \\
   \midrule
    WaveGen-S & 113M & 0.5k & 0.3M & 104.23&0.32&3.53  & 129.85 & 0.25 & 3.31 \\
    WaveGen-S & 113M & 0.5k & 0.5M & 26.33& 0.40& 3.55 & 92.88 & 0.27 & 3.22 \\
    WaveGen-S & 113M & 0.5k & 0.7M & 8.75&0.45&3.68  & 76.66 & 0.33  & 3.38 \\
    WaveGen-S & 113M & 0.5k & 0.9M &6.02&0.47&3.65 & 58.53 & 0.35 & 3.37 \\
    WaveGen-S & 113M & 0.5k & 1.1M & 4.96 & 0.47 & 3.72 & 56.58&0.36 & 3.36\\
   \midrule
    WaveGen-B & 407M & 0.5k & 0.3M & 28.19& 0.42& 3.60 & 93.01 & 0.31&3.33\\
    WaveGen-B & 407M & 0.5k & 0.5M & 9.19& 0.49& 3.60 & 76.53&0.33&3.20\\
    WaveGen-B & 407M & 0.5k & 0.7M & 4.32 & 0.51& 3.66 & 51.43& 0.39&3.32\\
    WaveGen-B & 407M & 0.5k & 0.9M & 3.69 & 0.52 & 3.65 & 44.82 & 0.39& 3.33\\
    WaveGen-B & 407M & 0.5k & 1.1M & 2.91 & 0.53& 3.67 & 41.56& 0.40 & 3.30 \\
   \midrule
    WaveGen-S & 113M & 120k & 0.3M & 91.23 &  0.32 & 3.27 & 148.32&0.21&3.02 \\
    WaveGen-S & 113M & 120k & 0.5M & 23.40 &  0.43 & 3.47  & 108.22& 0.29&3.17  \\
    WaveGen-S & 113M & 120k & 0.7M & 8.50 &  0.45 & 3.60  & 77.88&0.33&3.25 \\
    WaveGen-S & 113M & 120k & 0.9M & 5.00 &  0.48 & 3.59 & 69.43&0.34&3.25 \\
    WaveGen-S & 113M & 120k & 1.1M & 4.27 &  0.49 & 3.60 & 69.69& 0.35&3.28\\
   \midrule
    WaveGen-B& 407M & 120k & 0.3M & 63.51 & 0.44 & 3.47 & 143.51 & 0.29 & 3.15  \\
    WaveGen-B & 407M & 120k & 0.5M & 18.37&0.53&3.48  & 90.07&0.38 & 3.23\\
    WaveGen-B & 407M & 120k & 0.7M & 4.42&0.56&3.58 &58.73&0.40&3.22 \\
    WaveGen-B& 407M & 120k & 0.9M & 2.91& 0.57& 3.56 & 60.91&0.42 & 3.22\\
    WaveGen-B & 407M & 120k & 1.1M & 2.37& 0.59& 3.60 & 41.91&0.46&3.29  \\
    \bottomrule
  \end{tabular}
  }
\end{table}
\begin{table}[h]
  \caption{Additional Training Results}
  \label{table:additional}
  \centering
\resizebox{1\textwidth}{!}{
  \begin{tabular}{l|ccc|ccc|ccc}
    \toprule
    Model & Params. & Data &  \makecell{Training \\ Steps}   &  \makecell{\\ WER($\downarrow$)} &   \makecell{Basic CFG (2.0)\\SIM($\uparrow$)} &  \makecell{\\DNSMOS($\uparrow$)} &  \makecell{\\ WER($\downarrow$)} &   \makecell{No CFG\\SIM($\uparrow$)} &  \makecell{\\DNSMOS($\uparrow$)}    \\
   \midrule
    WaveGen-B & 407M & 120k & 1.1M & 2.37& 0.59& 3.60 & 41.91&0.46&3.29  \\
    WaveGen-B & 407M & 120k & 1.2M & 1.98&0.59&3.61& 39.67&0.46&3.31  \\
    WaveGen-B & 407M & 120k & 1.3M & 1.85&0.60&3.60& 36.67&0.46&3.30  \\
    \bottomrule
  \end{tabular}
  }
\end{table}
\begin{table}[h]
  \caption{CFG results}
  \label{table:cfg}
  \centering
\resizebox{0.8\textwidth}{!}{
  \begin{tabular}{l|cccc|ccc}
    \toprule
    Model & Params. & Data &  \makecell{Training \\ Steps} & CFG  &  \makecell{\\ WER($\downarrow$)} &   \makecell{Modified CFG (2.0)\\SIM($\uparrow$)} &  \makecell{\\DNSMOS($\uparrow$)}    \\
   \midrule
    WaveGen-B & 407M & 120k & 1.1M & 0.5 & 4.26&0.605&3.55 \\
    WaveGen-B & 407M & 120k & 1.1M & 1 & 2.61&0.625&3.60 \\
    WaveGen-B & 407M & 120k & 1.1M & 2 & 1.77&0.635&3.60 \\
    WaveGen-B & 407M & 120k & 1.1M & 3 & 1.93&0.637&3.52 \\
    WaveGen-B & 407M & 120k & 1.1M & 4 & 1.87&0.633&3.37 \\

    \bottomrule
  \end{tabular}
  }
\end{table}

\section{Additional results}
Table \ref{table:trainingsteps} shows the objective evaluation results of each model with respect to the number of training steps. The results indicate that model performance consistently improves as the number of training steps increases.

Due to limited computational resources, we trained the model only up to 1.1M steps, corresponding to four epochs on the Emilia dataset. Nevertheless, WaveGen-B continued to improve with additional training steps, as shown in Table \ref{table:additional}. These results suggest that end-to-end TTS models have strong potential for further improvement with larger-scale training.

Table \ref{table:cfg} shows the effect of CFG on the WaveGen-B model. 
Increasing the CFG scale substantially improves intelligibility, reducing WER from 4.26 to 1.77 when the CFG scale is increased from 0.5 to 2.0. 
Speaker similarity also improves consistently up to CFG 3.0, indicating that stronger conditioning helps the model better preserve the target speaker characteristics. 
However, excessively large CFG values slightly degrade perceptual quality, as reflected by the decrease in DNSMOS at CFG 3.0 and 4.0. 
These results show that CFG provides an effective trade-off between intelligibility, speaker similarity, and audio quality, with CFG 2.0 achieving the best overall balance in our experiments.
\newpage

\begin{figure}
    \centering
    {\includegraphics[width=1\textwidth]{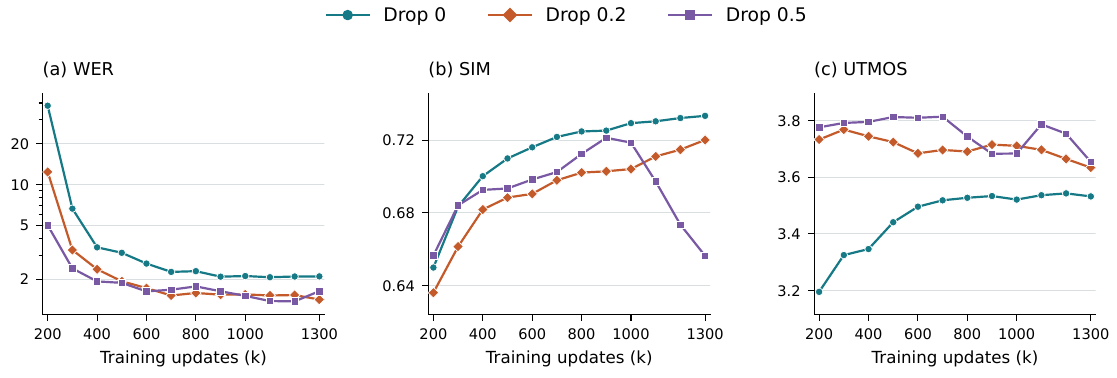}}   \vspace{-0.5cm}
    \caption{Spectral Dropout}
    \label{fig:dropout} \vspace{-0.4cm}
\end{figure} 

\section{Multilingual WaveGen} \label{sec:multi-lingual-wavegen} \vspace{-0.2cm}

We conducted multilingual text-to-speech comparison using Emilia (EN and ZH) datasets. For a fair comparison, we utilize the same architecture of F5-TTS to demonstrate the effectiveness of waveform-level diffusion training and Spectral Forcing. Then, we adopt the time scheduling methods of WavTTS to compare the effect of Spectral Forcing. For efficient waveform generation, we employ 50Hz resolutions using hop size of 480 for 24 kHz waveform diffusion. 

\subsection{Spectral Loss} \vspace{-0.2cm}
Basically, we introduce multi-scale spectral v-loss for our models. However, we adopt multi-scale Mel-spectrogram x-loss to compare Spectral Forcing based on WavTTS implementation. We observe that multi-scale Mel-spectrogram loss could improve the initial alignment learning and entire objective performances including WER and SIM. However, this loss induces buzzing sound on the generated waveform. To address this perceptual limitation, we released the perceptually-improved models together using multi-scale spectral v-loss which could significantly improve the perceptual quality. While it shows lower performance in objective metrics, we highly recommend adopting our new objectives for perceptual quality. We released the model finetuned with multi-scale spectral v-loss for perceptually improved speech synthesis.

\subsection{Spectral Dropout} \label{sec:spectral_dropout} \vspace{-0.2cm}
We further investigated the use of spectral dropout in spectral forcing to improve robustness to noise in multilingual datasets. Figure \ref{fig:dropout} presents the objective evaluation metrics for different spectral dropout rates. The results suggest that higher dropout rates can improve initial alignment when training on noisy waveforms, but may also cause the spectral path to collapse during training. Without spectral dropout, the model tends to rely on the noisy Mel-spectrogram for waveform reconstruction. We therefore selected a dropout rate of 0.2 to reduce this reliance while maintaining the stability of the spectral path.

\subsection{Sampling Steps} \vspace{-0.2cm}
With large-scale training, Table \ref{table:NFE_multi} shows that the model achieves better performance across all evaluated NFEs than that reported in Table \ref{table:NFE}. Furthermore, we plan to release an accelerated version of WaveGen that enables efficient speech synthesis in only a few sampling steps.

\begin{table}[h]

  \caption{Analysis of NFE (Seed-en, WaveGen)}
  \label{table:NFE_multi}
  \centering
  \resizebox{6.0cm}{!}{
    \begin{tabular}{l|c|ccc}
      \toprule
      Model & NFE & WER & SIM & UTMOS \\
      \midrule
                & 1  & 93.10 & 0.19 & 1.30 \\
                & 2  & 23.71 & 0.42 & 1.31 \\
                & 4  & 2.76  & 0.64 & 2.34 \\
                & 8  & 1.70  & 0.75 & 3.44 \\
      WaveGen   & 10 & 1.72  & 0.75 & 3.53 \\
                & 16 & 1.69  & 0.76 & 3.65 \\
                & 32 & 1.50  & 0.75 & 3.68 \\
                & 50 & 1.47  & 0.75 & 3.68 \\
      \bottomrule
    \end{tabular}
  }

\end{table}
 
\end{document}